\documentclass[
  journal=pasa,
  manuscript=research-paper, 
  year=2025,
  volume=37,
]{cup-journal}

\usepackage{microtype,siunitx,booktabs}

\usepackage{soul}
\usepackage[colorlinks=true,linkcolor=blue, citecolor=blue]{hyperref}
\usepackage{amsmath,amssymb, systeme}
\usepackage{subfigure}
\usepackage[normalem]{ulem} 
\usepackage{changes}

\defcitealias{YatesJones_2023}{Paper~I}
\newcommand{\PaperI}{\citetalias{YatesJones_2023}}

\defcitealias{Stewart_2025}{Paper~II}
\newcommand{\PaperII}{\citetalias{Stewart_2025}}

\title{CosmoDRAGoN III: Shaping the Afterlife -- How Progenitors and Environments Sculpt Radio Galaxy Remnants}

\author{Georgia S.C. Stewart}
\affiliation{School of Natural Sciences, University of Tasmania, Private Bag 37, Hobart, TAS, 7001, Australia}

\author{Stanislav S. Shabala}
\affiliation{School of Natural Sciences, University of Tasmania, Private Bag 37, Hobart, TAS, 7001, Australia}

\author{Patrick M. Yates-Jones}
\affiliation{School of Natural Sciences, University of Tasmania, Private Bag 37, Hobart, TAS, 7001, Australia}

\author{Ross J. Turner}
\affiliation{School of Natural Sciences, University of Tasmania, Private Bag 37, Hobart, TAS, 7001, Australia}

\author{Raffaella Morganti}
\affiliation{ASTRON, the Netherlands Institute for Radio Astronomy, Oude Hoogeveensedijk 4, 7991 PD, Dwingeloo, The Netherlands}
\alsoaffiliation{Kapteyn Astronomical Institute, University of Groningen, Postbus 800, 9700 AV Groningen, The Netherlands}

\author{Martin G. H. Krause}
\affiliation{Centre for Astrophysics Research, University of Hertfordshire, College Lane, Hatfield, Herts AL10 9AB, UK}

\author{O. Ivy Wong}
\affiliation{CSIRO Space and Astronomy, PO Box 1130, Bentley, WA 6102, Australia}
\alsoaffiliation{International Centre for Radio Astronomy Research, University of Western Australia, 35 Stirling Highway, Crawley, Western Australia 6009, Australia}

\author{Chris Power}
\affiliation{International Centre for Radio Astronomy Research, University of Western Australia, 35 Stirling Highway, Crawley, Western Australia 6009, Australia}
\alsoaffiliation{ARC Centre of Excellence for All Sky Astrophysics (ASTRO 3D), Australia}

\author{Martin J. Hardcastle}
\affiliation{Centre for Astrophysics Research, University of Hertfordshire, College Lane, Hatfield, Herts AL10 9AB, UK}

\doi{10.1017/pasa.2025.??}

\received {dd Mmm YYYY}
\revised  {dd Mmm YYYY}
\accepted {dd Mmm YYYY}
\published{dd Mmm 2025}

\keywords{active galactic nuclei, radio remnants, numerical hydrodynamic simulations} 

\begin{document}

\begin{abstract}

Identifying remnant radio-loud active galactic nuclei (AGNs) is challenging due to their diverse morphological and spectral characteristics. Using three-dimensional hydrodynamic simulations of 15 radio galaxies, we investigate how the spectral evolution of remnants depends on progenitor power, active lifetime, environment, and underlying dynamics. The simulations span low-density group and high-density cluster environments re-gridded from smooth-particle-hydrodynamic cosmological simulations. The resulting remnants exhibit a wide range of morphologies, from amorphous structures to double-lobed forms. We find that jet power correlates with the spectral slope. As the remnant lobes evolve, we find surface brightness depends strongly on environment: group remnants are systematically dimmer and more amorphous than cluster remnants, highlighting a potential observational bias against these low-surface-brightness sources. In our models, we estimate that the peak surface brightness of a low-redshift, 50 Myr-old remnant from a low-power progenitor in a 10$^{13}$ M$_{\odot}$ group environment should be routinely detectable at the 3$\sigma$ level with LOFAR, although 20–30\% of the emission would remain undetectable within a reasonable integration time. We find young remnants exhibit low-frequency ($150$–$1400$ MHz) spectral indices that overlap with active sources, and follow a consistent and established spectral-evolution sequence: significant curvature ($\alpha_{1400}^{6000} - \alpha_{150}^{1400} > 0.5$) develops before an ultra-steep low-frequency index ($\alpha_{150}^{1400} > 1.2$). The results presented in this work are intended as a reference point for current and upcoming low-frequency studies of radio remnants.
\\

\end{abstract}

\section{Introduction}
\label{sec:int}

\subsection{The Lifecycle of RLAGN Jets}
\label{subsec:AGN lifecycle} 

Many radio-loud active galactic nuclei (RLAGN) jets appear to operate on a duty cycle. Active phases in which pairs of relativistic, collimated jets are produced are followed by a remnant, or quiescent, phase where the lobes are left to radiate and fade away before the restart of jet production. The active stage of RLAGN evolution has been well studied observationally and theoretically since the pioneering observations of \cite{BOLTON_1949}, \cite{JENNISON_1953}, \cite{Mills_1957}, and seminal modelling work of \cite{REES_1971} and \cite{Scheuer_1974}. The remnant phase, however, is much less studied. \\

Accumulating large samples of remnant radio galaxies is necessary for a complete understanding of the entire RLAGN lifecycle. For example, \cite{Shabala_2020} argued that modelling a combination of active \textit{and} remnant sources is necessary to robustly constrain the lifetimes of RLAGN using the equations of \cite{Turner_2018}. More recently, \cite{quici_2024} constrained the jet lifetime function using a sample of only observed remnant candidates. Hydrodynamic simulations have shown that radio lobes can have an appreciable effect on the host environment even after the jets have ceased. Buoyantly rising remnant lobes can carry central gas to larger radii, drawing out filaments, reducing the central density of the ICM and accelerating the thermal exchange between lobe and ambient material \citep{Heath_2007,Revaz_2008,Werner_2010,Guo_2017,Chen_Heinz_Ensslin_2019, Zhang_2022, Husko_2023}. \cite{Pope_2010} also demonstrated AGN-blown bubbles could transport mass to higher radii, preventing cool material from accumulating in the central galaxy. \cite{Basson_2003}, \cite{Brighenti_2015}, and \cite{English_2019} used hydrodynamic simulations to show that once the jets become inactive and the lobes separate, off-axis cluster gas settles back into the gravitational potential. The energy stored in the rising lobes decreases rapidly as they continue to do work on their surroundings, eventually reaching pressure equilibrium with the environment.

\subsection{Identifying RLAGN Remnants}
\label{subsec:Remnant Selection Methods}

The identification of radio remnants is a challenging task, and often a combination of spectral and morphological data is required to confirm remnant candidates. Metrics for determining remnant status based on spectral properties depend on theoretical assumptions about the ageing process of synchrotron-emitting plasma. After the jet flow ceases, the radio lobes are no longer supplied with fresh, high-energy electrons. For a population of electrons with energies in the range $E$ to $E+dE$, the average timescale for energy losses via synchrotron processes is $$\tau = \frac{E}{dE/dt} = \frac{3m_{\rm{e}}c}{4\sigma_{\rm{T}}u_{\rm{B}}\gamma}$$ 
where $\sigma_{\rm{T}}$ is the Thomson cross-section, $u_{\rm{B}} = B^2/2\mu_{\rm{0}}$ is the magnetic field energy density, $c$ is the light speed, $m_{\rm{e}}$ is the electron mass, and $\gamma$ is the Lorentz factor. Since $dE/dt \propto -E^2$, the highest energy electrons will radiate faster than those at lower energies. Hence, the spectral index $\alpha$ which describes the slope of the source flux density $S$ with frequency $\nu$ such that $S_{\nu}\propto\nu^{-\alpha}$, is expected to steepen, starting at the higher frequency end and progressing to lower frequencies as the emitting plasma ages \citep{Jaffe_Perola_1973, Komissarov_Gubanov_1994}.\\

The majority of radio remnant identification studies use information from multi-frequency radio spectra (often in combination with morphological properties) to help classify remnant candidates. \cite{Brienza_2017}, \cite{Mahatma_2018} and \cite{Jurlin_2021} used a two-point spectral index threshold of $\alpha\geq 1.2$ between 150 MHz and 1400 MHz to identify aged plasma. These are typically classified as ultra-steep spectrum (USS) sources. A limitation of any low-frequency two-point spectral index cut is that young remnants, whose spectra have not yet steepened significantly, may be missed—an effect highlighted in modelling by \cite{Brienza_2016} and \cite{Godfrey_2017}. Moreover, candidate remnant samples presented in \cite{Brienza_2017}, \cite{Mahatma_2018}, and \cite{Jurlin_2021} showed a spectral index overlap with known active sources.
In addition to two-point spectral indices, the spectral curvature parameter (SPC) introduced by \cite{Murgia_2011} has also been used. It is defined as SPC $= \alpha_{\rm{high}} - \alpha_{\rm{low}}$, where SPC $> 0.5$ indicates aged plasma. Because spectral steepening typically occurs first at high frequencies, SPC may help identify younger remnants whose low-frequency spectra remain relatively flat. However, if the spectrum is already steep at both high and low frequencies, SPC may offer limited additional insight and could result in the misclassification of older sources.\\

Morphological criteria have been used to infer remnant status either on their own \citep[e.g.][]{Saripalli_2012} or, more often, as a complement to spectral information \citep[e.g.][]{Brienza_2016, Mahatma_2018}. Low, diffuse surface brightness (e.g. SB$_{150 \rm{ MHz}} < 50 \rm{mJy / arcmin^2}$, \citealp{Brienza_2017}) and amorphous structures are identified with remnant lobes as they undergo adiabatic as well as radiative losses. The timescales over which these characteristics develop are likely dependent on environment and progenitor properties. For example, if the source remains overpressured at the end of the active phase or if it propagates into a low-density environment, adiabatic losses will be more rapid \citep[][]{Kaiser_Cotter_2002, Perucho_2014, English_2019, Stewart_2025}. In addition, the absence of compact features (i.e. a bright pc-scale core and hotspots) associated with active radio sources is often taken as an indicator of inactivity \citep{Giovannini_1988}. These features are expected to disappear on timescales similar to the light travel time through the remnant lobe. \cite{Mahatma_2018} used the absence of a core to search for radio remnants in the Herschel-ATLAS field. They found that 11 of 33 candidate remnants had no core emission at 6 GHz, indicating true remnant lobes. The spectral index values of those 11 sources varied between $1.5 \geq \alpha_{150}^{1400} \geq 0.5$. In \cite{Quici_2021}, the authors identified 10 candidate remnant sources based on a lack of core emission at 5.5 GHz. Three out of 10 showed steep spectra and low surface brightness, while seven out of ten showed compact regions in the lobes at 5.5 GHz. In \cite{Jurlin_2021}, the authors also identified USS sources with core emission at 6 GHz. Later, in \cite{quici_2024}, the authors found that 39 out of a sample of 79 remnant candidates show no core emission. These findings suggest that the interpretation of compact features in remnant candidates can be challenging or that many identified remnant candidates may not be true remnants.\\

\subsection{This Work}

Spectral and morphological metrics have traditionally been used to identify remnant candidates, however most of these are based on broad expectations from theory rather than detailed models. We use three-dimensional hydrodynamic simulations of active and remnant radio galaxies to trace the evolution of commonly used remnant selection metrics such as surface brightness, spectral index, and spectral curvature. We determine how these metrics depend on progenitor and environmental properties. \\

This is the third work in the  Cosmological Double Radio AGN (CosmoDRAGoN) project; a suite of three-dimensional hydrodynamic simulations of AGN jets run in environments derived from smooth particle hydrodynamics (SPH) cosmological galaxy formation simulations of galaxy clusters and groups \citep[][hereafter \PaperI]{YatesJones_2023}. It is also the second paper in the CosmoDRAGoN series to explicitly examine the active-to-remnant transition and the subsequent evolution of AGN remnants in detail. In the preceding companion work \citep[][hereafter \PaperII]{Stewart_2025}, we looked at the mechanisms that drive cocoon and shock evolution during the active and remnant phases, and compared the trends seen in the simulations with analytical modelling. In the present study, we generate synthetic surface brightness images and model the spectral evolution using post-processing techniques to generate synthetic radio spectra across a range of frequencies and redshifts. Throughout the paper, we assume the standard $\Lambda$CDM cosmology with the following parameters: $\Omega_{\rm{M}} = 0.307$, $\Omega_{\rm{B}} = 0.048$, $\Omega_{{\Lambda}} = 0.693$, $h = 0.678$ \citep{Plank_Collab_2016}. This cosmology is consistent with the simulation catalogue from which our environments have been derived. \\

The paper is structured as follows: In Section \ref{sec:sim_setup}, we describe the numerical setup before we detail the post-processing pipeline used to generate synthetic radio emission in Section \ref{post-processing}. In Section \ref{sec:lowzremnants} we probe the morphological and spectral evolution of our active and remnant radio sources at low ($z=0.05$) redshift to establish a baseline for their emission properties before extending the analysis to higher redshifts. In Section \ref{sec:higherzremnants}, we look at the emission properties out to $z=1$; a redshift below which most remnant candidates are identified. We summarise our results relating to the commonly used remnant selection metrics in Section \ref{sec:remnant selection timescales} and remark on the environment-dependence of remnant evolution in Section \ref{sec:Discussion}. We conclude in Section \ref{sec:conclusions}.


\section{Numerical Setup and Initial Conditions}
\label{sec:sim_setup}

Our simulation setup is described in the preceding works in this series: \PaperI ~  and \PaperII.  As the present work is a continuation of the dynamical study presented in ~\PaperII, the simulation suite is identical to that work. We briefly summarise the key features here and refer the reader to those papers for further details. Our three-dimensional hydrodynamic simulations of active and remnant radio jets are taken from the \mbox{CosmoDRAGoN} simulation suite and have been run using a modified version of the \textsc{pluto} 4.3 code \footnote{http://plutocode.ph.unito.it/}. For those simulations where the initial jet injection has a velocity of 0.98$c$ (see Table \ref{tab:sim_props}) we include relativistic physics.\\

The jets are injected as a conical, bipolar pair of outflows with half-opening angle $\theta_{\rm{j}}$ from within a spherical region of radius $r_o=0.75$, centred at the grid origin. They are described initially as a fluid with velocity $v_{\rm{j}}$, density $\rho_{\rm{j}}$, pressure $P_{\rm{j}}$, adiabatic index $\Gamma$, and kinetic power $Q_{\rm{j}}$. In the relativistic regime, the one-sided jet power is related to the initial pressure, density, velocity and cross-sectional area of the jet head, $A_{\rm{j}}$ by

\begin{equation}\label{Eq. jet power}
Q_{\rm{j}} = \left[\frac{\Gamma}{\Gamma - 1} P_{\rm{j}} \gamma^2 + \gamma(\gamma-1)\rho_{\rm{j}} c^2   \right]v_{\rm{j}} A_{\rm{j}}
\end{equation}

\noindent for speed of light, $c$, and  bulk flow Lorentz factor, \\$\gamma = 1/\sqrt{1-v_{\rm{j}}^2/c^2}$. For non-relativistic flows, Eq. \ref{Eq. jet power} reduces to $\frac{1}{2}\rho_{\rm{j}} v_{\rm{j}}^3A_{\rm{j}}$. \\

For the purposes of generating synthetic surface brightness projections in post-processing (see Section \ref{post-processing}), our simulations use \textsc{pluto}'s inbuilt particle module to inject 16 Lagrangian tracer particles every 0.01 Myr \footnote{This number and frequency of particle injection was found to accurately sample the volume of the lobe for the range of total source sizes considered in this work.}. Each particle is regarded as a macro-particle that represents a packet of relativistic electrons. These are injected within the same injection region as the jet and follow the streamlines of the jet fluid. \\

The remnant phase begins when the jet injection onto the grid is terminated. At this point, the injection region is no longer defined on the grid, and the cells that were within that region take on similar density, pressure, and velocity values to the surrounding cells. Candidate radio remnants have been observed to have total linear sizes in the range of several hundred kpc \citep[e.g][]{Brienza_2016} to a few tens of kpc \citep[e.g.][]{Singh_2021}. To capture this variety, we terminate each set of injection parameters (i.e. the parent model) at three points: when the total source size has reached $180$ kpc, $60$ kpc, and $20$ kpc. Each source is evolved for approximately 100 Myr past the time of switch-off, $t_{\rm{off}}$, or until it approaches the edge of the grid.\\

All simulation runs are carried out on a three-dimensional Cartesian grid with maximum physical dimensions (X, Y, Z) = (400, 400, 400) kpc. In each dimension, five grid patches are used. These comprise one central uniform grid with a resolution of 0.05 kpc/cell, an inner stretched grid ($\pm2.5$ kpc, $\pm 10$ kpc) with a stretching ratio of $\sim1.0076$, and an outer stretched grid ($\pm10$ kpc, $\pm 200$ kpc) with a  stretching ratio of $\sim1.0084$. This gives grid resolutions of 0.05 kpc/cell at $\pm2.5$ kpc, 0.08 kpc/cell at $\pm10$ kpc and 0.84 kpc/cell at $\pm100$ kpc. Outflow boundary conditions are imposed on all grid edges to prevent non-physical reflections at the extremities. \\

In keeping with \PaperII, we propagate our simulated jets through realistic environments derived from SPH cosmological simulations. In particular, we derive our environments from \textsc{the three hundred project} \citep{Cui_2018} catalogue of re-simulated galaxy clusters with full-physics hydrodynamics (including cooling, star formation, black hole growth, and feedback). We have selected suitable halos 0003 and 0031 \footnote{assessing the suitability of the halos involves checking they are hydrostatically stable with no evidence of merger events at the epoch of interest; $z=0$ in this work} from galaxy cluster simulation 002 to represent a cluster and a group environment, respectively. These have halo masses of $2.0 \times 10^{14}M_{\odot}$ for the cluster and $1.9 \times 10^{13}M_{\odot}$ for the group. Environments 0003 (cluster) and 0031 (group) are re-gridded onto a three-dimensional mesh grid suitable for use in \textsc{pluto} by interpolating the cluster quantities (particle density, pressure, momentum density and force
density) using the following smoothing function, 
\begin{equation}
    A_s(r) \approx \Sigma_i m_i \frac{A_i}{\rho_i}W(|r-r_i|, h_i)
\end{equation}
where $m_i$ and $\rho_i$ are the mass and density of the particle, respectively, $h_i$ is the particle smoothing length, and $W(r, h) = M_4(r)/h^3$ is a smoothing function dependent on radius and smoothing length. For further details on the derivation of the environments from \textsc{the three hundred project}, the interested reader is referred to ~\PaperI.\\

The raw outputs of the simulation are the \textsc{pluto} grid and particle files. The grid files are output typically every $1-2$ Myr and contain the fluid density, pressure, and three-dimensional velocity. The particle files contain the identification numbers of the Lagrangian macro-particles, their densities, pressures, injection times, time of last shock acceleration, and three-dimensional position and velocity information.\\

Each simulated jet is initialised with a jet kinetic power, fluid velocity, and half-opening angle. We consider two jet powers ($10^{38}$ W and $10^{36}$ W) and two initial velocities ($0.98$ c and $0.01$ c). High jet kinetic powers, such as the $10^{38}$ W power used here, and strongly relativistic velocities are consistent with Fanaroff-Riley Type II (FR-II) dynamics \citep{Antognini_2012, Perucho_2021} while low jet powers and sub-relativistic velocities are consistent with FR-I sources \citep{Laing_2014}. \\

With five sets of initial conditions (the initial jet power, velocity, half-opening angle, and two cosmological environments) and three termination points per set, we obtain a suite of 15 simulations. In the discussions that follow, each simulation is designated a run code of the form \textbf{Qaa-vbb-$\theta$cc-Xdd} for initial jet power, \textbf{Qaa} (in log Watts), velocity, \textbf{vbb} (as a fraction of $c$), half-opening angle $\theta$ (degrees), environment (\textbf{X$=$G} for group and \textbf{X$=$C} for cluster), and total physical size at termination in kpc, \textbf{dd}. We show all simulation parameters in Table \ref{tab:sim_props}. \\

\begin{table}[h!]
\centering
\begin{threeparttable}
\caption{Parameters for the simulation runs. $Q_{\rm{j}}$ is the one-sided kinetic jet power in Watts, $v_{\rm{j}}$ is the initial jet velocity as a fraction of the speed of light, $\theta_{\rm{j}}$ is the half-opening angle in degrees, $t_{\rm{on}}$ s the duration of the active phase in Myr, and $t_{\rm{on}}+t_{\rm{off}}$ is the total runtime of the simulation. The run code for each model is given in the last column. }
\label{tab:sim_props}
\begin{tabular}{ c c c c c c c } \toprule
\hline Env. & $Q_{\rm{j}}$ & $v_{\rm{j}}$ & $\theta_{\rm{j}}$  & $t_{\rm{on}}$ & $t_{\rm{on}} + t_{\rm{off}}$ & Run Code\\
 & (W) &  (c) & (deg) & (Myr) & (Myr)& \\ 
\hline 
\text {Cluster} &    $10^{36}$ & 0.01 & 25   & 5 &  100 & Q36-v01-a25-C20\\
                &             &      &      & 22 & 120 &Q36-v01-a25-C60\\
                &              &      &      & 100 & 200 &Q36-v01-a25-C180\\
\hline
                &    $10^{38}$ & 0.98 & 25   & 1.4 &  100& Q38-v98-a25-C20\\
                &              &      &      & 7.5 & 110 &Q38-v98-a25-C60\\
                &              &      &      & 56  & 155 &Q38-v98-a25-C180\\
\hline
                &    $10^{38}$ & 0.98 & 7.5  & 0.5 & 100 &Q38-v98-a7.5-C20\\
                &              &      &      & 1.5 & 95 &Q38-v98-a7.5-C60\\
                &              &      &      & 15  & 110 &Q38-v98-a7.5-C180\\
\toprule
\text {Group}   &    $10^{36}$ & 0.01 & 25   & 6 & 110 &Q36-v01-a25-G20\\
                &            &        &      & 16 & 115 &Q36-v01-a25-G60\\
                &            &      &        & 100 & 200 &Q36-v01-a25-G180\\
\hline
                &    $10^{38}$ & 0.98 & 25   & 0.8 & 75 &Q38-v98-a25-G20\\
                &            &      &       &  3.8 & 70 &Q38-v98-a25-G60\\
                &            &      &       &  22.8 & 95& Q38-v98-a25-G180\\
 \hline
\bottomrule
\end{tabular}
\end{threeparttable}
\end{table}

The simulations of the active phase of jet evolution were carried out using the \textit{Gadi} facility provided by the National Computational Infrastructure, Australia. Simulations of the remnant phase of evolution were carried out using the \textit{kunanyi} facility provided by the Tasmanian Partnership for Advanced Computing. The maximum CPU time of a simulation was approximately 790 000 CPU hours.


\section{Post Processing}
\label{post-processing}

\subsection{Approach}
\label{ref:approach}

We model the synchrotron emission from our simulations using a post-processing framework that couples hydrodynamic simulations with an analytical treatment of radiative losses. This follows the methods of \cite{Yates_Jones_2022, YatesJones_2023}, which builds on the iterative analytical formulation of \cite{Turner_2018b_resolved_spectral_evolution} and the synchrotron equations explained in \cite{Longair_2010}. The main calculation steps are summarised in Section~\ref{subsec:radio_emission_calcs}. \\

Briefly, classical dynamical models \citep[e.g.][]{KDA_1997} and subsequent numerical studies show that, on large scales, the magnetic field in radio lobes broadly follows the thermal pressure, reproducing key observational features such as spectral steepening with distance from the hotspot \citep{Alexander_1987}. \cite{Turner_2018b_resolved_spectral_evolution} demonstrated that the dynamics of the backflow are important for a quantitative interpretation of spectral ageing. By combining hydrodynamic backflows with analytical post-processing, they showed that particle mixing can account for much of the dynamical-spectral age discrepancy, wherein spectral ages appear younger than dynamical ages by factors of several. Recently, \cite{Jerrim_2025} further confirmed the importance of particle mixing and the validity of post-processing synchrotron emission under the assumption of a constant ratio between the magnetic field energy density and pressure (the same approach as used in the present work, see Section \ref{subsec:radio_emission_calcs}). Comparing three-dimensional hydrodynamic and magnetohydrodynamic simulations of jets and environments similar to those used here, they found that while magnetic fields affect fine-scale surface brightness, the integrated lobe flux densities and spectra remain consistent between the hydrodynamic and magnetohydrodynamic models to within tens of per cent. This hydrodynamic approach enables efficient exploration of an expansive parameter space in jet power and environment across a grid of 15 simulations, and allows emission predictions at multiple redshifts with minimal additional computational cost.\\

\subsection{Calculating the Radio Emission}
\label{subsec:radio_emission_calcs}

We refer the interested reader to \cite{Yates_Jones_2022} and \PaperI~ for comprehensive details on the emission calculations, but we provide a summary of the three key steps relevant to the present study below. 

\begin{itemize}
    \item \textbf{Particle Shock Flagging}. \textsc{pluto}'s Lagrangian particle module allows for the detection of shocks. Following the shock capturing process of \cite{Mignone_2012}, macro-particles are flagged as being shocked if the velocity divergence is negative, $\nabla \cdot \vec{v} < 0$, and the gradient of the pressure at the location of the particle exceeds a defined dimensionless threshold $\epsilon_{\rm{p}}$ (equation B1 in \citealp{Mignone_2012}). Multiple thresholds can be used, but in this work, we use a single shock threshold of $\epsilon_p = 5$ corresponding to a Mach number of $\mathcal{M}\sim 2.24$, which is consistent with \PaperI.
    \item \textbf{Emissivity}. The process for generating synthetic radio maps from the simulated radio sources is calculated following the PRAiSE implementation in \cite{Yates_Jones_2022}. The radio emission is assumed to come from populations of high-energy, synchrotron-emitting particles. The emissivity per unit volume and solid angle is calculated in the rest-frame $\nu'$ of the electron packet following 
    \begin{multline*}
        j'_{\nu '} = \frac{K_{\rm{s}}}{4\pi}(\nu')^{\frac{1-s}{2}}\frac{\eta^{\frac{s+1}{4}}}{(\eta + 1)^{\frac{s+5}{4}}} \\ \times p(t)^{\frac{s+5}{4}}\left[\frac{p(t_{\rm{acc}})}{p(t)} \right]^{1-\frac{4}{3\Gamma_{\rm{c}}}}\left(\frac{\gamma_{\rm{acc}}}{\gamma} \right)^{2-s}.
    \end{multline*}

    \noindent The emissivity depends on the ratio of the electron to magnetic field energy densities, $\eta$, the exponent $s$ of the assumed power-law energy distribution of electrons when they are shock accelerated $N(E, t)dE = N_{\rm{0}}E^{-s}dE$, the adiabatic index of the cocoon $\Gamma_{\rm{c}}$, and the Lorentz factors and particle pressures at the time of last acceleration, ($\gamma_{\rm{acc}}, p(t_{\rm{acc}})$) and current time ($\gamma, p(t)$). $K_{\rm{s}}$ is a radio source specific constant (see Section 2.2 in \citealp{Yates_Jones_2022}) with the form,

    \begin{multline*}
        K(s) = \frac{\kappa(s)}{m_e^{(s+3)/2}c(s+1)}\left[ \frac{e^2\mu_0}{2(\Gamma_c -1)}\right]^{(s+5)/4}  \left(\frac{3}{\pi}\right)^{s/2} \\ \times \left[\frac{\gamma_{\rm{min}}^{2-s}-\gamma_{\rm{max}}^{2-s}}{s-2} - \frac{\gamma_{\rm{min}}^{1-s}-\gamma_{\rm{max}}^{1-s}}{1-2}\right]^{-1},
    \end{multline*}

    which depends on the minimum and maximum Lorentz factor limits ($\gamma_{\rm{min}}$, $\gamma_{\rm{max}}$), electron mass $m_e$, speed of light $c$, and the vacuum permeability $\mu_0$. We take an injection index of $s=2.2$ corresponding to a spectral index of $\alpha = 0.6$, minimum and maximum Lorentz factors $\gamma_{\rm{min}} = 500$ \citep{Godfrey_2009} and $\gamma_{\rm{max}} = 10^5$ \citep{KDA_1997, Hardcastle_Krause_2013, Turner_2015}. The lobe adiabatic index is calculated within the simulation according to the Taub-Matthews equation of state \citep{Taub_1948, Mathews_1971}. The constant $\kappa(s)$ is
    \begin{equation}
        \kappa(s) = \frac{\Gamma(\frac{s}{4} + \frac{19}{12})\Gamma(\frac{s}{4} - \frac{1}{12})\Gamma(\frac{s}{4} + \frac{5}{4})}{\Gamma(\frac{s}{4} + \frac{7}{4})}.
    \end{equation}

    \noindent Since magnetic field strength is not calculated in the simulation, we require a way to map the magnetic field strength from the hydrodynamical quantities. Following \cite{Kaiser_Alexander_1997}, the pressure in the radio lobes is a function the electron, magnetic field, and thermal energy densities, $p = (\Gamma_c -1)(u_{\rm{e}} + u_{\rm{B}} + u_{\rm{th}})$ where $u_{\rm{th}}$ is assumed to be negligible. The magnetic field strength can then be derived by taking an equipartition $\eta = \frac{u_B}{u_e}$ factor between the magnetic and particle energy densities,
    \begin{equation}
        B = \left( \frac{2\mu_0 p}{\Gamma_c - 1 } \frac{\eta}{\eta+1}\right)^{1/2}.
    \end{equation} 
    The value of the equipartition factor used in this work is 0.03, representative of moderate power radio sources \citep{Croston_2014, Ineson_2017}.
    
    \item \textbf{Projection}. The final product of our post-processing pipeline is a two-dimensional surface brightness grid obtained from a line-of-sight projection. In this work, we consider only plane-of-the-sky projections, and so the line of sight is taken to be perpendicular to the simulation grid and projection grid.
\end{itemize}

\section{Spectral Properties at Redshift z=0.05}
\label{sec:lowzremnants}

In this section, we investigate the source evolution for all simulations at a redshift of 0.05. In Section \ref{sec:higherzremnants} we extend this analysis to a redshift of 1. All surface brightness data is obtained using the post-processing PRAiSE method outlined in Section \ref{post-processing} and calculated for an input redshift of 0.05. We first consider the morphological and spectral characteristics of our active and remnant simulations in Section \ref{subsec:morphology} and relate these to features observed in real radio sources. In Section \ref{subsec:spectra}, we consider the evolution of spectral properties from the active to remnant stage. In particular, we consider the time evolution of surface brightness, spectral index and spectral curvature. We convolve our surface brightness projections with a circular Gaussian of full-width-half-maximum 5 arcseconds. This is similar to the \mbox{LOFAR} 6 arcsecond beam size used by \cite{Mahatma_2018}, \cite{Jurlin_2020, Jurlin_2021} and \cite{Singh_2021} for radio emission at 150 MHz and the \mbox{FIRST} 5 arcsecond beam for radio emission at 1400 MHz \citep[e.g.][]{Mahatma_2018, Jurlin_2021}. In the following subsections, we assume infinite sensitivity. In this approach, a source is only `undetectable' when the integrated line-of-sight emissivity from the Lagrangian macro-particles is zero.

\subsection{Morphology} 
\label{subsec:morphology}

\begin{figure*}
    \centering
    \includegraphics[width=0.8\linewidth]{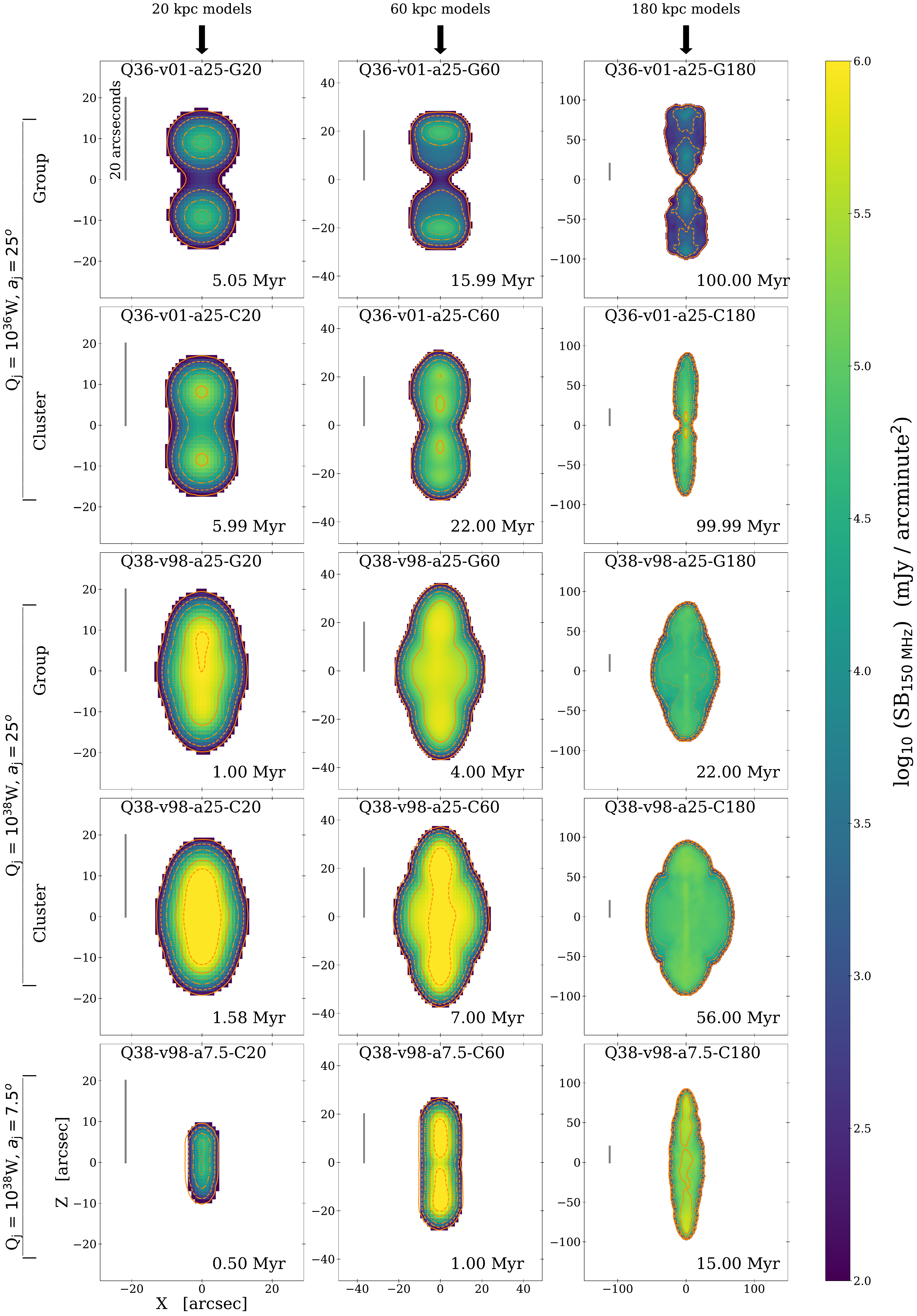}
    \caption{The spatially resolved surface brightness maps at 150 MHz for all active simulations in mJy arcminute$^{-2}$. The snapshots are taken at the last grid output before $t_{\rm{on}}$. The total simulation time is indicated in the lower right corner of each panel. The orange contours show 150 MHz surface brightness data at 10$^2$, 10$^3$, 10$^4$, 4$\times$10$^4$, 2$\times$10$^5$, and 10$^6$ mJy arcminute$^{-2}$. The physical dimensions are the same for colour maps in any given column. The vertical grey line in each panel indicates an angular size of 20 arcseconds.}
    \label{fig:gridactives}
\end{figure*}

\begin{figure*}
    \centering
    \includegraphics[width=0.8\linewidth]{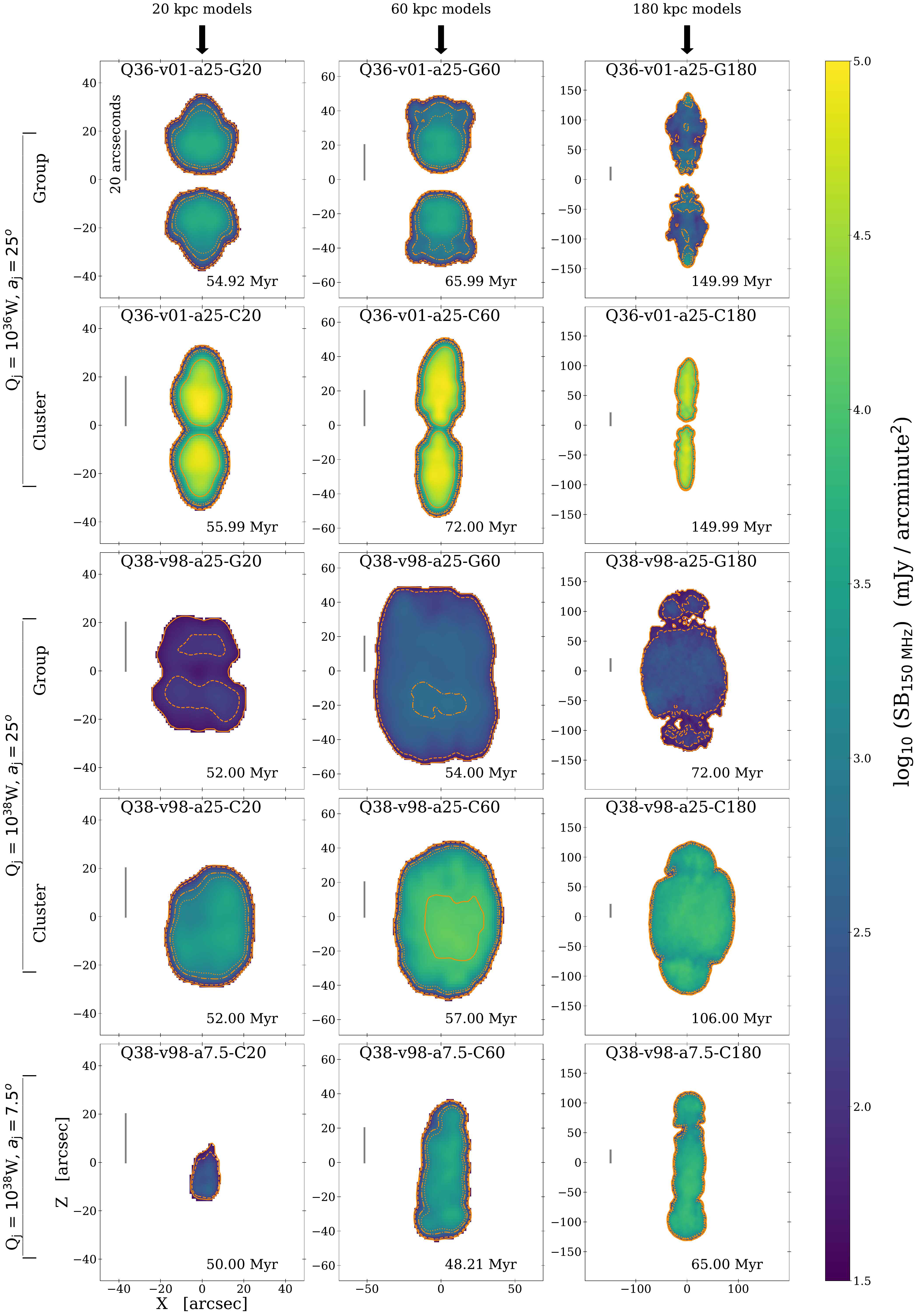}
    \caption{An analogous plot to Fig \ref{fig:gridactives} but showing each simulation in the remnant phase at $t_{\rm{off}} +  50$ Myr and with 150 MHz surface brightness contours at 45, 100, 500, 10$^3$, and 10$^4$ mJy arcminute$^{-2}$.}
    \label{fig:gridremnants}
\end{figure*}

In this section, we first briefly consider the key morphological features of our active sources in Section \ref{subsec:morphology:actives}  before examining the morphological changes in the remnant phase in Section \ref{subsec:morphology:remnants}. In Figs. \ref{fig:gridactives} and \ref{fig:gridremnants} we show low-frequency (150 MHz) surface brightness maps for all simulations at the end of the active phase (Fig. \ref{fig:gridactives}) and 50 Myr into the remnant phase  (Fig. \ref{fig:gridremnants}).

\subsubsection{Active Sources}
\label{subsec:morphology:actives}

In the active phase, our simulated radio sources show varying radio morphologies as a result of their initial jet and environment conditions. Several features captured in our simulations are qualitatively similar to the key morphological characteristics of observed radio sources. The low-jet power, large cluster simulation (right-most panel in row two of Fig. \ref{fig:gridactives}) shows edge-darkened surface brightness profiles for the largest source sizes consistent with an FR-I morphology \citep[e.g.][]{Laing_2014}. For the same source in the group environment (right-most panel in row one of Fig. \ref{fig:gridactives}), the regions of brightest emission are near the head of the lobe despite its low, $Q_{\rm{j}} = 10^{36}$ W, and slow $v_{\rm{j}} = 0.01$ c, initial jet power and velocity. This appears similar to the subpopulation of low-luminosity FR-II radio sources (that sit below the canonical FR-I/FR-II divide of $L_{150} \sim 10^{26}$ W Hz$^{-1}$) identified by \cite{Mingo_2019} using the LOFAR Two-Metre Sky Survey (LoTSS). Those authors show that the low-luminosity FR-II subpopulation preferentially reside in low-mass environments, supporting the idea that low-power jets may show FR-II-like characteristics in sufficiently low-density environments. For smaller low-powered sources, the emission peaks are closer to the centre of the lobe. \\

The bottom three rows of Fig \ref{fig:gridactives} show all sources with high jet powers ($Q_{\rm{j}} = 10^{38}$ W) and strongly relativistic injection velocities ($v_{\rm{j}} = 0.98$ c). For these sources, the lobes are filled by backflow from a terminal shock which is strong enough to re-accelerate the Lagrangian macro-particles near the jet head. This yields surface brightness profiles that appear consistent with the edge-brightened morphologies of FR-II sources. The emission peaks at 150 MHz at the end of the lobes resemble the `hotspots' observed in these powerful radio galaxies \citep[e.g.][]{Hardcastle_1998}. These are particularly well observed in the 60 and 180 kpc switch-off jets (bottom three panels in the middle and right columns of Fig. \ref{fig:gridactives}, respectively). For the jets that are more compact, high-powered, and have wide half-opening angles (24 degrees, left panels in the third and fourth row), the emission is brightest towards the centre of the lobe along the jet axis.\\

\subsubsection{Remnant Sources}
\label{subsec:morphology:remnants}

Our remnant radio sources are characterised by comparatively (to the active phase) more uniform surface brightness distributions, reflecting the fading of the compact emission features described in Section \ref{subsec:morphology:actives}. \textit{Observed} remnant candidates are shown to have amorphous structures at low radio frequencies (e.g. `blob1' \citealp{Brienza_2016}, J102905+585721 \citealp{Jurlin_2021}, and J130532.49+315639.0 \citealp{Mahatma_2018}). A comparison between the corresponding panels in Fig. \ref{fig:gridremnants} and Fig. \ref{fig:gridactives} shows that, qualitatively, the remnants of compact, high-powered progenitors experience more rapid changes in their morphology than low-powered sources. Further to this, comparing equivalent sources across the group and cluster environments suggests that remnants in the group environment become distorted more rapidly than those in the cluster environment. This is consistent with the results of \PaperII ~ which show that the overpressure in group simulations is greater than cluster simulations at switch-off. \\

\subsection{Surface Brightness and Radio Spectra} 
\label{subsec:spectra}

We now consider how various spectral quantities change from the active (Section \ref{subsec:spectra:actives}) to remnant phase (Section \ref{subsec:spectra:remnants}). In the literature, attempts to classify remnant radio galaxies often take the surface brightness (for example, at low, 150 MHz, frequencies as in \citealp{Mahatma_2018, Jurlin_2021} and \citealp{Dutta_2022}), spectral index, and spectral curvature into consideration. We refer the reader back to Section \ref{subsec:Remnant Selection Methods} for details on the spectral selection metrics. We apply these same metrics to our simulated sources in this section.

\subsubsection{Active Sources}
\label{subsec:spectra:actives}

\begin{figure*}
    \centering
    \includegraphics[width=0.85\linewidth]{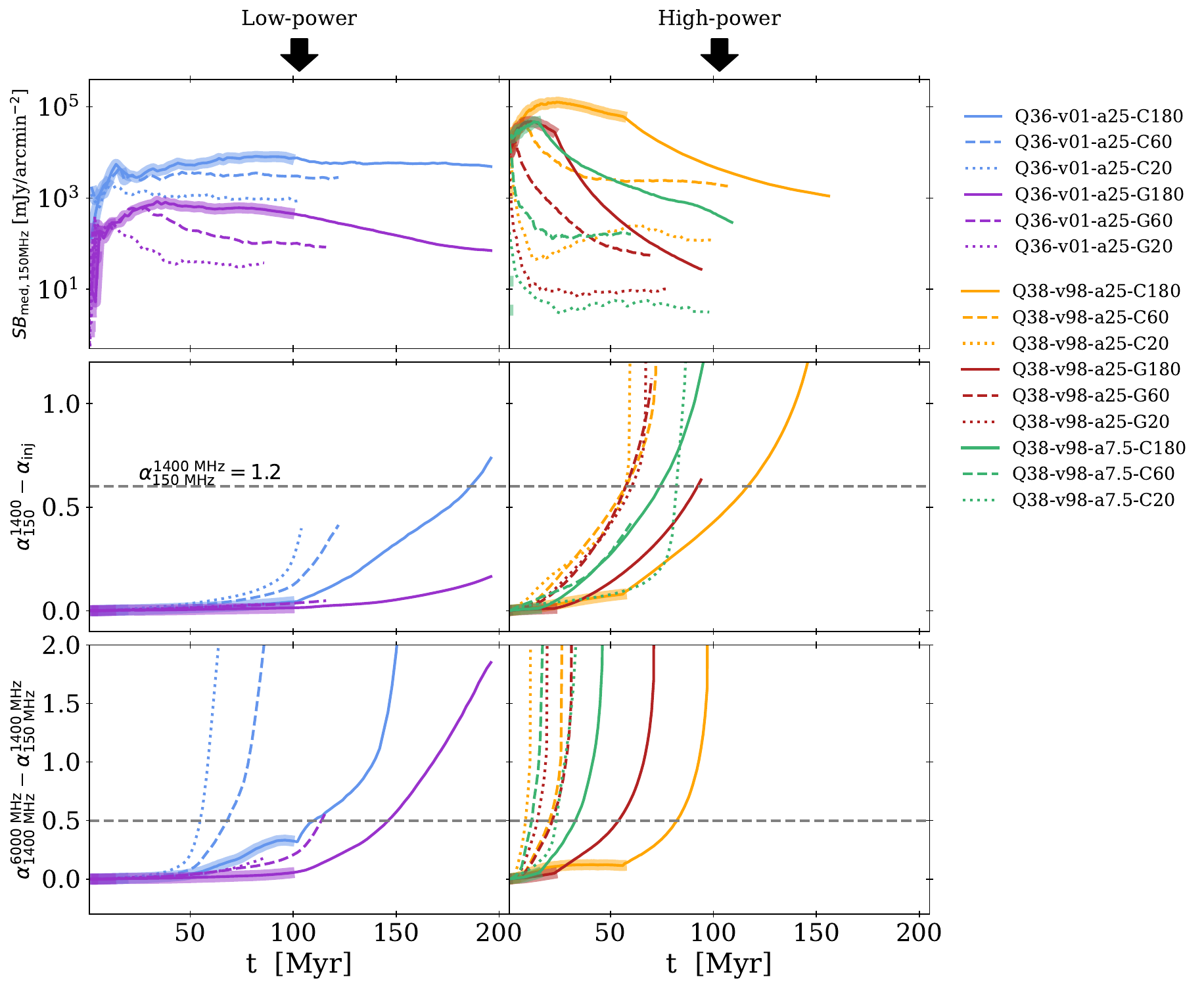}
    \caption{Temporal evolution of several spectral properties of our simulations. From top to bottom, we show: the median surface brightness at 150 MHz, the integrated two-point spectral index taken between 1400 MHz and 150 MHz, and the integrated spectral curvature between $\alpha_{1400}^{6000}$ and $\alpha_{150}^{1400}$. The active and remnant phases are denoted respectively by thick and thin lines. Tracks of the same colour indicate the same remnant progenitor properties, while the dotted, dashed and solid line styles denote the 20, 60, and 180 kpc switch-off points, respectively. The simulations with low kinetic jet powers ($10^{36}$ W) are shown in the left column, and those with high powers ($10^{38}$ W) are on the right.}
    \label{fig:z=0.05spectralpropertiesgrid}
\end{figure*}

We show the time evolution of several spectral properties for each simulation in Fig. \ref{fig:z=0.05spectralpropertiesgrid}. In the top row of Fig. \ref{fig:z=0.05spectralpropertiesgrid}, we show the median surface brightness at 150 MHz as a function of time. In order to compare our simulations to other recent works \citep[e.g][]{Brienza_2017, Quici_2021}{}{}, we give the surface brightness in units of mJy arcminute$^{-2}$ at a redshift of 0.05. A key result is that simulations in the low-density group environment are characteristically dimmer than their cluster counterparts throughout the active phase (denoted by the thicker lines). At $t_{\rm{on}}$, the median 150 MHz surface brightness of all group jets is almost an order of magnitude lower than equivalent simulations in the cluster environment. This agrees with earlier numerical \citep[e.g.][]{Hardcastle_Krause_2013, English_2019} and analytic \citep[e.g.][]{Turner_2022} modelling results. In our group simulations, reduced surface brightness compared to cluster simulations is due to the lower density of radiating material. While both have similar active times, low-powered group jets inflate larger lobes, resulting in greater adiabatic losses. In high-powered jets, group runs reach the target source length in about half the time of equivalent cluster jets while producing lobes of similar volume. \\

\begin{figure}[h]
    \centering
    \includegraphics[width=\linewidth]{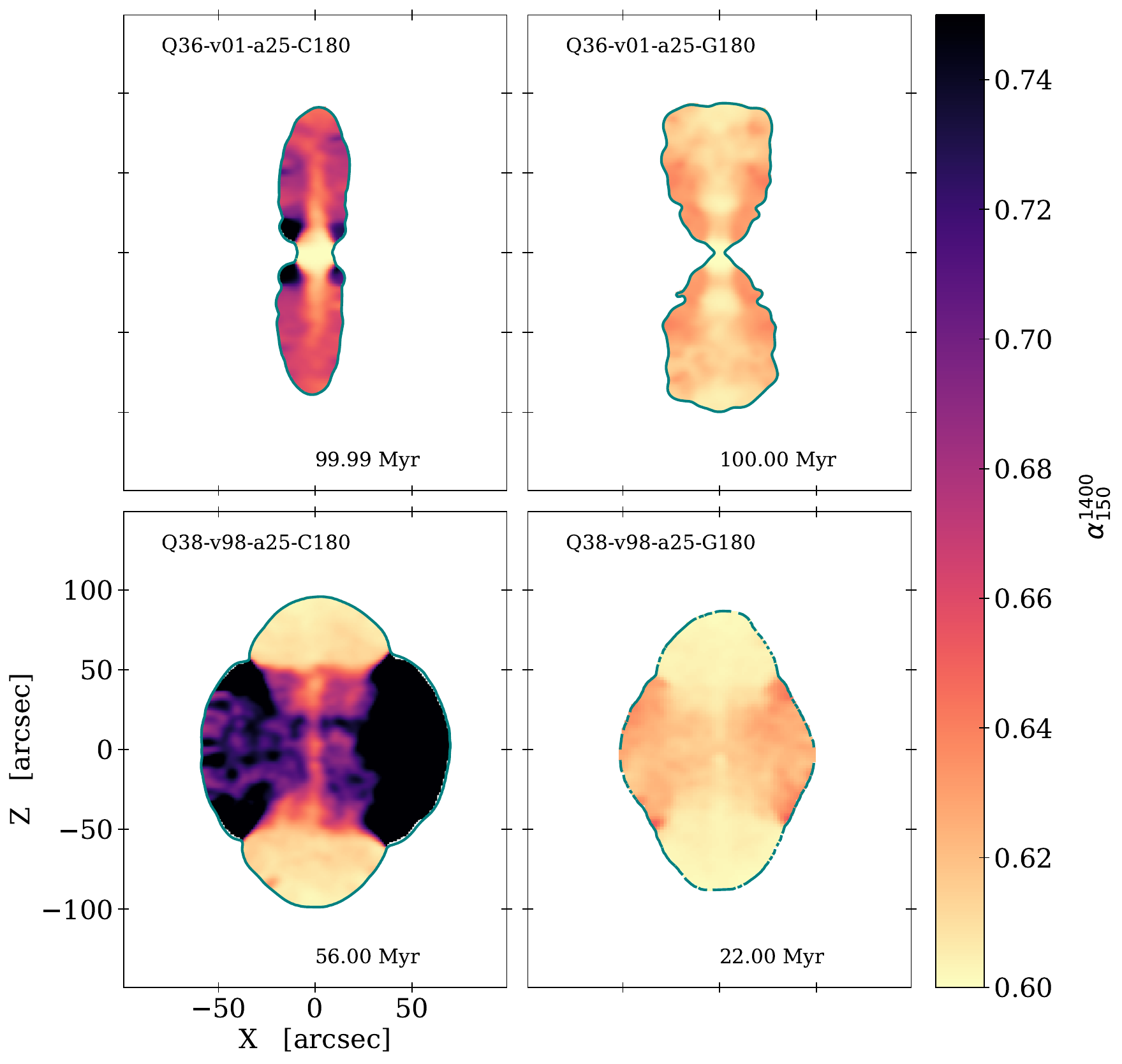}
    \caption{Spatial distributions of $\alpha_{150}^{1400}$ for a subset of our largest simulations during the active phase at the last grid output before $t_{\rm{on}}$ (exact times are indicated in the lower right). In the top row, we show the low-powered sources in the cluster (left) and group (right). In the bottom row, we show the high-powered wide simulations in the cluster (left) and group (right). Simulation codes are shown in the top left. The contour outlines the low-frequency, 150 MHz radio emission at 0.05 mJy arcsecond$^{-2}$.}
    \label{fig:some_active_spectral_index_colormaps}
\end{figure}

We show the spatial distribution of the spectral index, $\alpha_{150}^{1400}$, for a subset of our active sources in Fig. \ref{fig:some_active_spectral_index_colormaps}. For high-powered sources (bottom row of Fig. \ref{fig:some_active_spectral_index_colormaps}), the most spectrally aged lobe material (dark purple) resides at the lateral extremities, where the oldest backflow is continuously pushed out by younger jet material flowing back from the terminal shock. For low-powered sources (top row of Fig. \ref{fig:some_active_spectral_index_colormaps}), the lobe spectral indices steepen closer to the injection region.\\

In the second row of Fig. \ref{fig:z=0.05spectralpropertiesgrid}, we consider the change in integrated spectral index between 150 MHz and 1400 MHz with respect to the injection index $\alpha_{\rm{inj}}$. At each frequency $\nu$, the spectral index is calculated as $\alpha_\nu = \frac{\partial \log S_\nu}{\partial\log \nu}$ for total flux density $S_\nu$.  We remind the reader that we take $\alpha_{\rm{inj}} = 0.6$ in this work. The frequency range from 150 MHz to 1400 MHz is  commonly used to compute the spectral index in recent works \citep[e.g.][]{Jurlin_2021}.  We have also indicated on Fig. \ref{fig:z=0.05spectralpropertiesgrid} where the integrated spectral index would be classified as ultra steep ($\alpha_{150}^{1400} = 1.2$; the dashed grey line). At $z=0.05$, no active sources would be considered ultra steep between 150 and 1400 MHz. This is unsurprising given the continuous injection of fresh radiating material. All extended sources (the sources with the longest active times) show a narrow range of $\Delta \alpha_{150}^{1400}$ from $0$ and $0.12$. There is a slight difference between group and cluster simulations, with cluster simulations having marginally steeper spectra (around $\Delta \alpha_{150}^{1400} \approx 0.03$).The difference between the two environments arises due to a longer radiating timescale for simulations in the group environment, where the assumed lobe magnetic field is lower. For more compact sources with shorter active times, the range of spectral indices is even narrower in both environments. \\

We consider the spectral curvature across 150 MHz, 1400 MHz, and 6000 MHz in the third row of Fig. \ref{fig:z=0.05spectralpropertiesgrid}. The spectral curvature is measured as the difference $\alpha_{1400}^{6000} - \alpha_{150}^{1400}$. We indicate a spectral curvature of $0.5$ with the grey dashed line; this denotes significant curvature typical of aged plasma \citep{Murgia_2011, Jurlin_2021}. Although we do not find any simulations with significantly curved spectrum during the active phase (our large, low-powered jet in the cluster environment Q36-v01-a25-C180 does come close with a spectral curvature of around 0.4 at switch-off), we find that the spectral curvature is consistently larger in the cluster, with low-powered cluster sources experiencing the greatest losses and having the greatest spectral curvature in the active phase.

\subsubsection{Remnant Sources}
\label{subsec:spectra:remnants}

We now consider the spectral evolution of our remnant sources. The rate of surface brightness fading is shown in the top row of Fig. \ref{fig:z=0.05spectralpropertiesgrid}. A comparison of the slopes of the plots in the left (low-power sources) and right (high-power sources) panels of Fig. \ref{fig:z=0.05spectralpropertiesgrid} shows that the fading rate increases with jet power. This is consistent with the results of \cite{Turner_2018}. Further to this, simulations run in the group environment fade more rapidly at 150 MHz than their equivalent cluster counterpart (compare the slopes of the red and orange lines in the top right panel of Fig. \ref{fig:z=0.05spectralpropertiesgrid}). For high-powered sources, the overpressure in the lobe coupled with the low density of the environment (see \PaperII ~) drives faster adiabatic expansion and shorter radiative loss timescales. The opening angle also has some impact on the rate of surface brightness fading. Immediately after the jets switch off, simulations with narrow opening angles dim rapidly compared to those with wide opening angles due to their smaller volumes (compare the orange and green lines, top right panel of Fig. \ref{fig:z=0.05spectralpropertiesgrid}). At later times, the fading rates become similar.\\

\begin{figure}
    \centering
    \includegraphics[width=\linewidth]{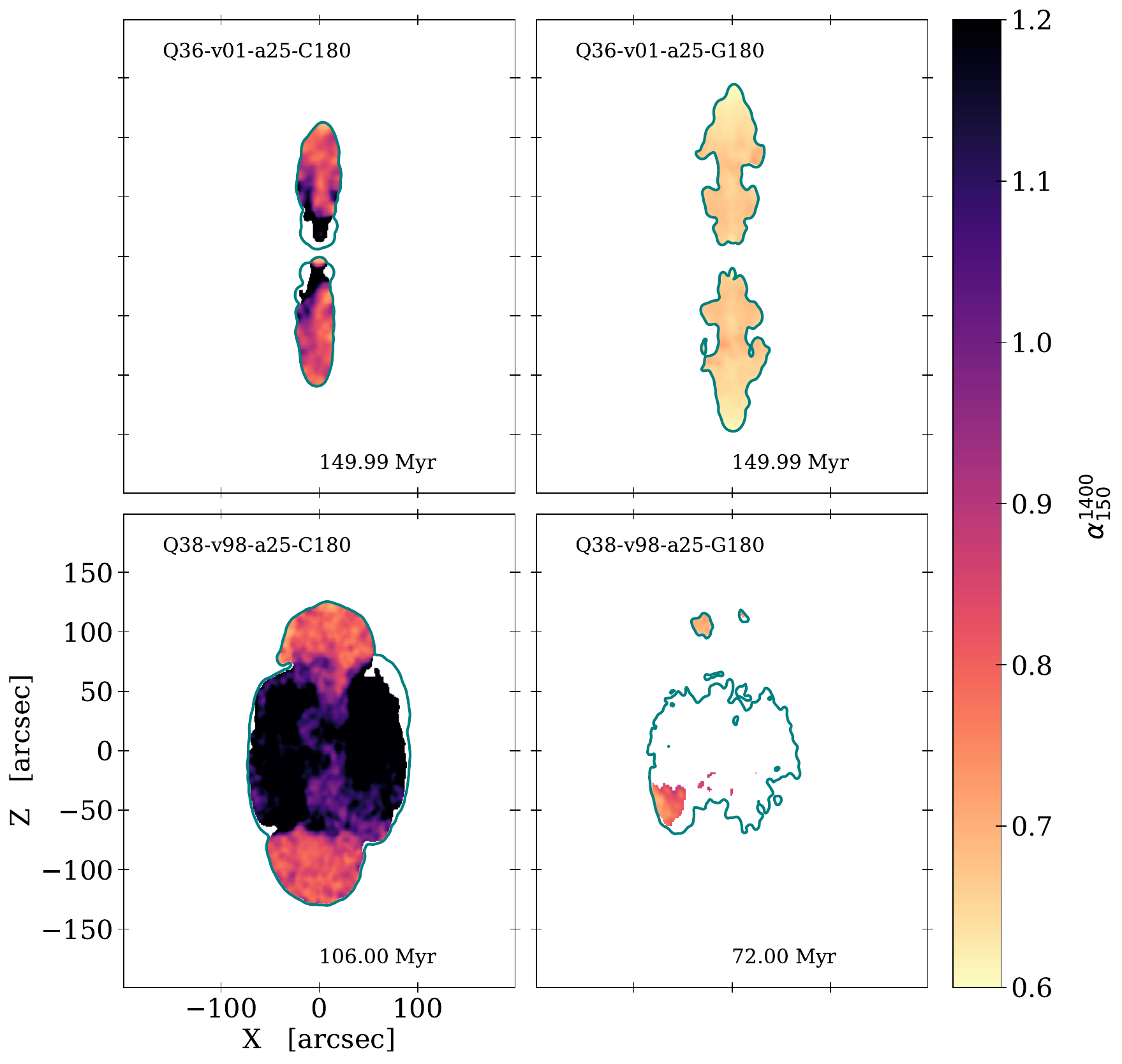}
    \caption{Analogous to Fig. \ref{fig:some_active_spectral_index_colormaps} but showing the spatial distribution of spectral indices 50 Myr into the remnant phase.}
\label{fig:some_remnant_spectral_indices_large_sources}
\end{figure}

In \PaperII ~, we showed that small, low-powered remnants continue to behave dynamically as if they were still active for a period of time after the jets switch off. This can occur over timescales similar to the previous active phase. In the top left panel of Fig. \ref{fig:z=0.05spectralpropertiesgrid}, the delay is also evident in the surface brightness at 150 MHz, where the remnant phase (thin lines) continues to track the active phase (thick lines) for an additional 10–20 Myr.\\

In the second row of Fig \ref{fig:z=0.05spectralpropertiesgrid}, we show the evolution of the spectral index into the remnant phase. At low redshift, only one low-powered source reaches an ultra-steep spectrum within the simulated time. Here, we define an ultra-steep spectrum as a change in spectral index of $\Delta \alpha = 0.6$, which denotes a steepening to $\alpha = 1.2$ from an injection spectral index of $0.6$. For high-powered sources, significant spectral steepening beyond $\Delta\alpha = 0.6$ occurs only during the remnant phase, after a period comparable to the duration of the preceding active phase. The rate of spectral steepening is faster in the cluster environment due to the older particle population. A narrower jet opening angle also appears to increase the rate of spectral steepening in remnants from high-powered sources (compare the orange and green lines in the second row of Fig. \ref{fig:z=0.05spectralpropertiesgrid}). The rate of spectral curvature steepening is also faster for remnants in the cluster environment, as shown by the third row of Fig. \ref{fig:z=0.05spectralpropertiesgrid}. All sources cross the spectral curvature $0.5$ threshold within their simulated time and before the spectra become steep between 150 MHz and 1400 MHz (compare rows two and three in Fig. \ref{fig:z=0.05spectralpropertiesgrid}). \\

The spatial distribution of spectra does not change appreciably between the active and early (first 50 Myr) remnant phases, as seen by comparing Figs. \ref{fig:some_active_spectral_index_colormaps} and \ref{fig:some_remnant_spectral_indices_large_sources}. In the top left-hand panels of Fig. \ref{fig:some_active_spectral_index_colormaps} and \ref{fig:some_remnant_spectral_indices_large_sources}, we see the most aged spectra remains at the base of the lobe, while the flatter spectral indices are found towards the top of the lobe and in the old jet channel. These general features are similar for simulation Q38-v98-a25-C180 in the lower right panel, where the distinct pockets of flat spectra at the head of the lobe remain in the remnant phase. In the case of simulation Q38-v98-a25-G180, the emission at 1400 MHz has faded considerably, such that the integrated spectral index is derived from a very small portion of the remnant lobe.

\section{Spectral Properties from Redshift 0.05 $< z <$ 1}
\label{sec:higherzremnants}

\begin{figure*}
    \centering
    \includegraphics[width=0.90\linewidth]{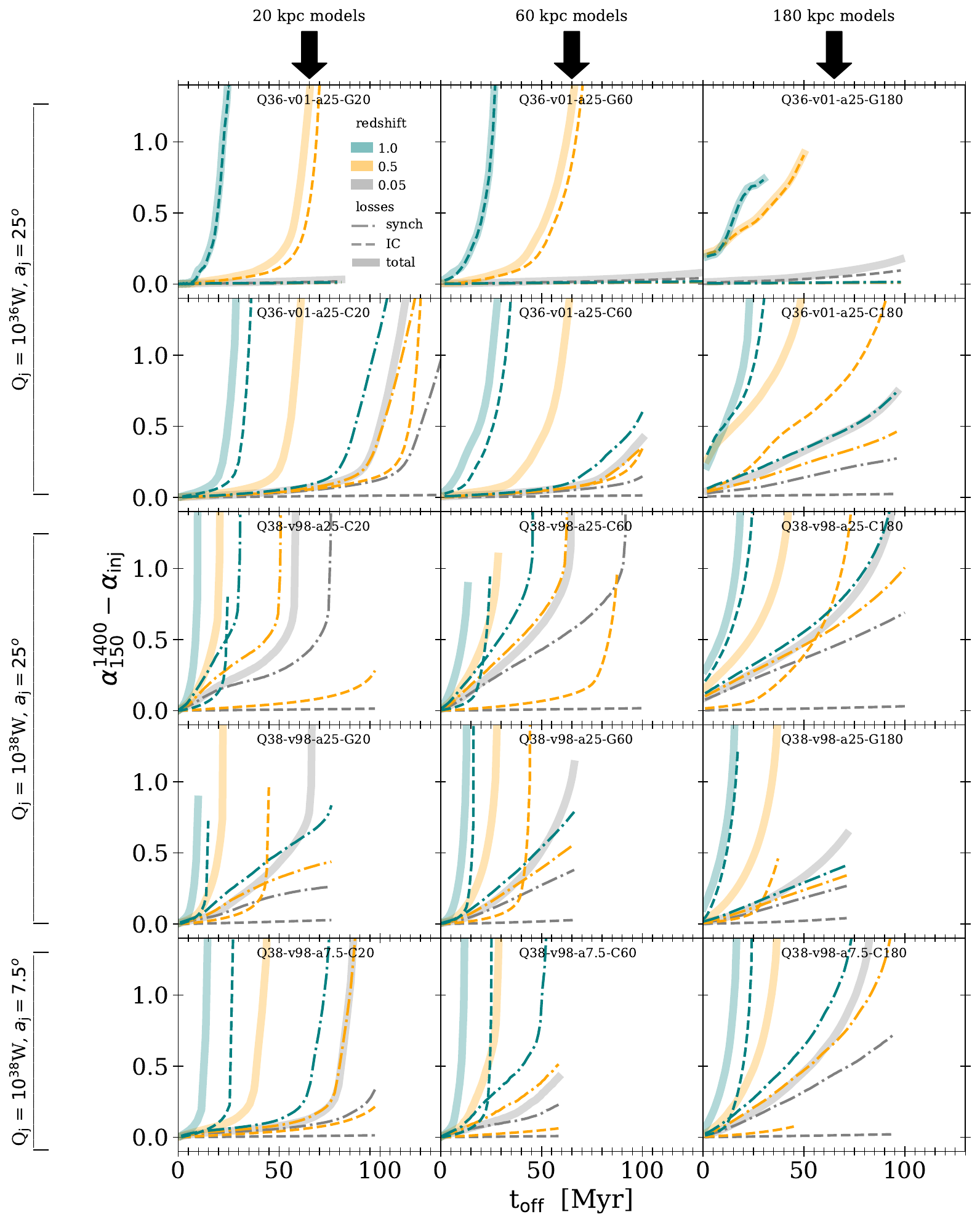}
    \caption{The evolution of the change in integrated spectral index (with respect to the injection index) in the remnant phase for inverse-Compton scattering (dashed lines), synchrotron (dot-dashed lines) and full losses (solid thick line). We include tracks at three different redshifts. $z=0.05$ is plotted in grey, $z=0.5$ in orange, and $z=1$ in green. The proximity of dot-dashed or dashed lines to the thick line indicates the dominance of synchrotron or inverse-Compton loss processes, respectively. The layout of this plot is analogous to Figs. \ref{fig:gridactives} and \ref{fig:gridremnants} such that each panel shows a single simulation, with low-powered progenitors in the top two rows and high-powered progenitors in the bottom three rows. Simulation identification codes are shown in the top middle of every panel.}
    \label{fig:loss_grid}
\end{figure*}

In this section, we probe the spectral properties of our simulation suite for redshifts $>0.05$. Recent low-frequency observations have collected candidate remnant samples out to spectroscopic redshifts of around $0.8$ and photometric redshifts $>1$ \citep[e.g.][]{Mahatma_2018, Jurlin_2021, Quici_2021}. In this work, we consider the source emission evolution out to $z=1$. At higher redshifts, radiative losses from the inverse-Compton scattering of cosmic microwave background (CMB) photons increase due to the higher density of the CMB. In Section \ref{subsec:loss_processes}, we analyse the contributions of synchrotron and inverse-Compton processes separately to the steepening of our simulated remnants for $ 0.05<z<1$. In  Section \ref{subsec:synthetic_populations}, we analyse the spectral properties of the entire simulated active and remnant population before looking at how the spatial distributions of surface brightness and spectral index change with increasing redshift in Section \ref{subsec:spatial_distributions}. Finally, in Section \ref{sec:core_prom} we describe the evolution of core prominence during the remnant phase.

\subsection{Loss Processes}
\label{subsec:loss_processes}

Using the methodology outlined in Section \ref{post-processing}, we are able to isolate the contributions of different loss mechanisms to the total spectral steepening. In particular, we separate the contributions from inverse-Compton scattering and synchrotron losses to explore the mechanisms that drive spectral steepening through a redshift range of 0.05 to 1. Adiabatic losses are included with all radiative loss processes. \\

In Fig. \ref{fig:loss_grid} we plot the temporal evolution during the remnant phase of the change in integrated spectral index value between $150$ MHz and $1400$ MHz ($\Delta\alpha_{150}^{1400}$) for three redshifts: $0.05$, $0.5$ and $1$. In this figure, we separate the contributions from synchrotron and inverse-Compton scattering processes in addition to plotting the total spectral index evolution (thick solid lines). The number of macro-particles in a simulation is fixed at jet switch-off. To ensure the remnant lobes are well-sampled, we only calculate the spectral index while at least 10\% of this macro-particle population are emitting.\\

In the case of a low-power jet in a low-density group environment where the magnetic field is small (top panel of Fig. \ref{fig:loss_grid}), the spectral steepening is almost entirely governed by inverse-Compton (plus adiabatic) processes. This occurs regardless of redshift and means that, at low redshift, the spectral index remains similar to the injection index. For a higher jet power in the same group environment (second row from the bottom), the magnetic field is higher, and synchrotron processes are dominant at low redshift. For all sources in the cluster environment, synchrotron processes are the dominant radiative loss mechanism for all sources at $z = 0.05$. During the first 40 Myr in the remnant phase, we find source spectral steepening to be very well described by synchrotron processes. After this time, the break frequency due to inverse-Compton scattering has shifted towards low frequencies, resulting in a non-negligible contribution to the overall spectral steepening. \\

At a redshift of $0.5$, the contributions from inverse-Compton scattering are naturally increased in all cases with the higher energy density of the CMB. In the cluster environment, synchrotron losses still remain dominant  for most high-jet power simulations, while for high-jet powers in group environments, synchrotron and inverse-Compton processes are comparable.\\

At a redshift of $1.0$, spectral steepening due to inverse-Compton scattering dominates all low-powered sources, resulting in similar spectral index values for a given source size regardless of environment. Similar conclusions can be drawn from the high-powered sources, with all sources experiencing similar, rapid spectral steepening after switch-off. For some high-powered simulations in the cluster environment, synchrotron loss processes are still dominant for a short period. \\

\subsection{Population Distributions of Radio Emission}
\label{subsec:synthetic_populations}

\begin{figure*}
    \centering
    \includegraphics[width=0.9\linewidth]{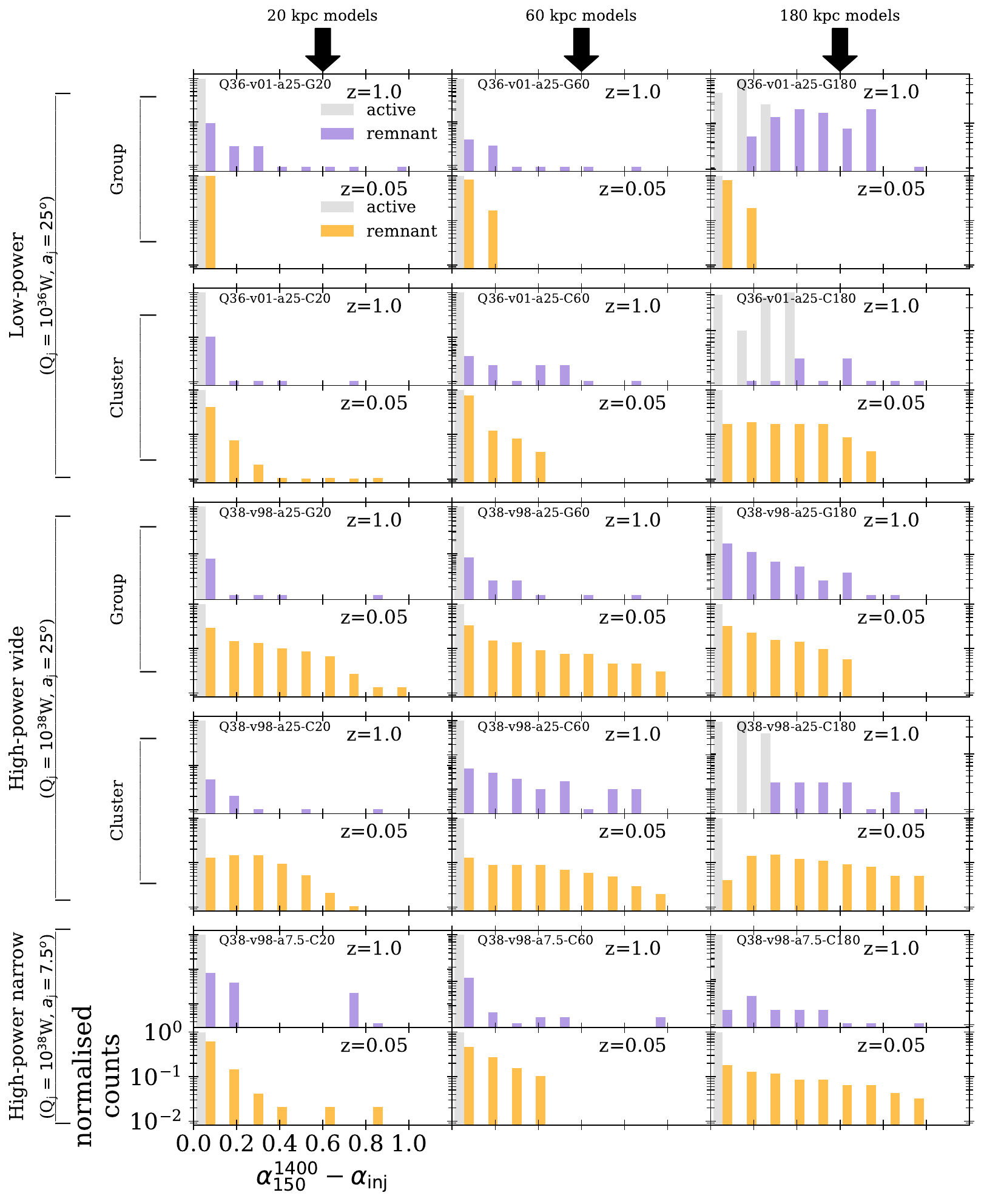}
    \caption{Histograms of the change in spectral index (relative to the injected spectral index) between 1400 MHz and 150 MHz for all simulations at z = 0.05 (orange) and z=1.0 (purple).  Active sources are shown in grey while remnants are colored by their redshift. Each simulation is sampled for the total number of available outputs and at 1 Myr intervals. The number of counts in the active and remnant phases has been time-weighted and normalised, representing the fraction of time spent in each spectral-index bin.}
    \label{fig:specindex_grid_all_sims}
\end{figure*}

We now consider the evolution of spectral index and spectral curvature for the entire simulated population across the active and remnant phases. Again, we consider sources between redshifts 0.05 and 1. In Fig. \ref{fig:specindex_grid_all_sims} we show $\Delta\alpha_{150}^{1400}$ histograms for all sources at $z=0.05$ (orange) and $z=1$ (purple). Each simulation is sampled at 1 Myr intervals for the total duration of the simulated time. For both redshifts, we confine the histograms to a range of $\Delta \alpha_{150}^{1400}$ from $0$ (corresponding to the injection spectral index of $0.6$) to $1.4$ (corresponding to a spectral index of $2$). There are some outliers which reach $\alpha_{150}^{1400} > 2$; however, at that time, there is very little emission present on the grid. At the two sampled redshifts ($z = 0.05, 1.0$), the majority of remnant objects have $\Delta \alpha_{150}^{1400} < 0.5$ and overlap with the active source population (grey solid bars and grey hashed bars). 
Our simulated population does not contain any source with an ultra steep spectrum ($\Delta\alpha_{150}^{1400} > 0.6$) at any point during the active phase at either redshift. The sources with the steepest spectral indices during the active phase are attributed to the 180 kpc switch-off simulations (specifically, Q36-a25-v01-G180, Q36-a25-v01-C180, and Q38-a25-v98-C180), which have the longest active times (and the oldest particles) for their combination of jet and environment properties (see the far right column of Fig. \ref{fig:specindex_grid_all_sims}). Except for these simulations, all other remnants in our suite require timescales longer than the previous active phase to steepen above  $\Delta\alpha_{150}^{1400} > 0.6$. This is consistent with the idea that remnant selection using an ultra-steep, low-frequency spectral index will be biased towards the most aged plasma \citep[][]{Godfrey_2017, Brienza_2017}. \\

\begin{figure*}
    \centering
    \includegraphics[width=0.9\linewidth]{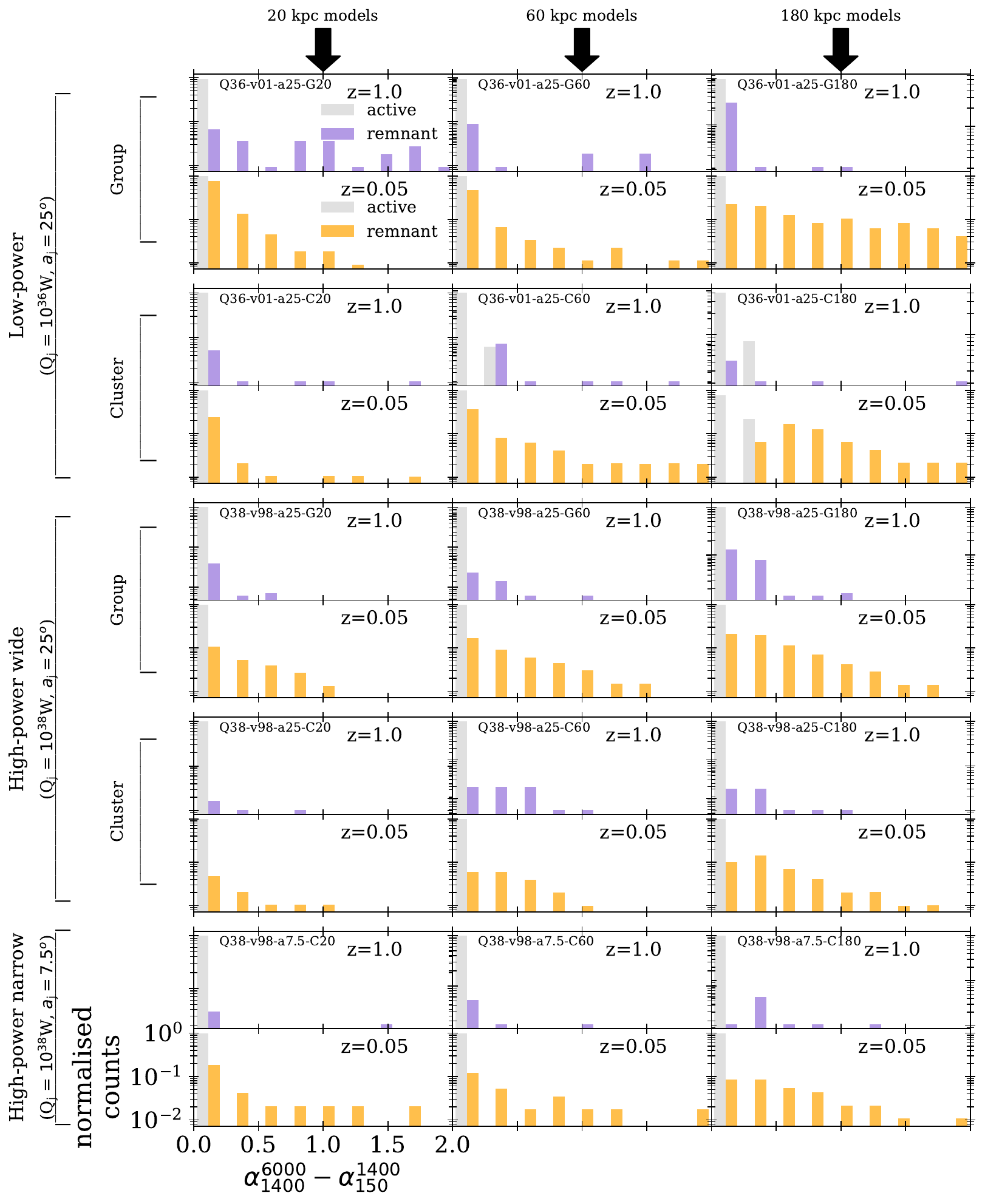}
    \caption{Histograms of the spectral curvature ($\alpha_{1400}^{6000} - \alpha_{150}^{1400}$) for all simulations at z = 0.05 (orange) and z=1.0 (purple). This figure is analogous in layout to Fig. \ref{fig:specindex_grid_all_sims}.}
    \label{fig:speccurve_grid_all_sims}
\end{figure*}

We present a similar grid of histograms that show the spectral curvature counts for all simulations in Fig. \ref{fig:speccurve_grid_all_sims}. We calculate the spectral curvature as $\alpha_{1400}^{6000} - \alpha_{150}^{1400}$ where the surface brightness projections at all three frequencies are beam-matched at 5 arcseconds. In all histograms, the low remnant counts at $z=1.0$ are due to the lack of emission at high frequencies. Once the jet switches off, the $6000$ MHz emission fades rapidly at high redshifts. For some simulations (e.g. the high-power, compact simulations Q38-v98-a25-G20 and Q38-v98-a25-C20), this timescale is less than 10 Myr. Our simulated population does not capture any active sources that exceed a spectral curvature value of 0.5. Generally, the highest density of counts for active and remnant sources is $< 0.5$ with the tail of the distribution in the direction of higher values of spectral curvature. This suggests that once curvature starts to develop in the spectra of remnant sources, there is not a long period of time before fading at higher frequencies limits the detectability of the source at $1400$ and $6000$ MHz. This agrees with the modelling of \cite{Godfrey_2017} and \cite{Hardcastle_2018}.

\subsection{Spatial Distributions of Radio Spectra}
\label{subsec:spatial_distributions}

In this section, we consider the spatial distribution of surface brightness and $\alpha_{150}^{1400}$ over the redshift range 0.5 to 1 for a sample of our simulations. Our surface brightness data are again obtained using the PRAiSE framework with the desired input redshift. Some observed remnant candidates are shown to have spatially resolved spectral index maps that exhibit a wide range of values across the source. An example of this is B2 0924+30 \citep{Shulevski_2017} with $ 0.7 \lesssim \alpha_{150}^{1400} \lesssim 1.6$. On the other hand, other remnant candidates show relatively uniform spectral indices across most of the source (for example, see Fig. 3 in \citealp{Morganti_2021_LOFAR_Apertif_lifecycles}). \\

\begin{figure*}
    \centering
    \includegraphics[width=0.90\linewidth]{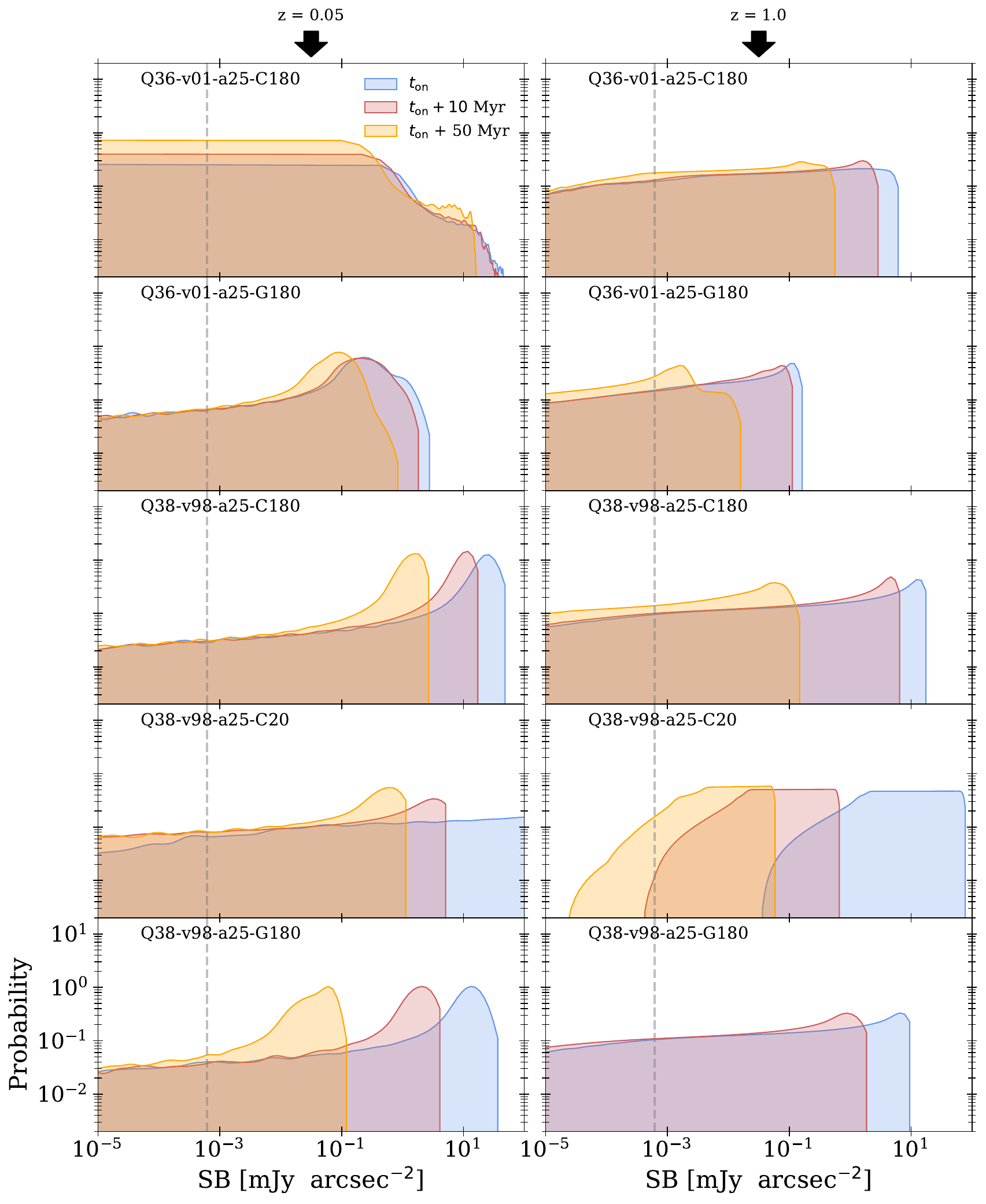}
    \caption{Probability density plots of surface brightness for five representative simulations for $z=0.05$ (left column) and $z=1.0$ (right column), showing how the range of surface brightness decreases for remnant sources. The vertical grey line in each panel indicates an instrument sensitivity of 23 $\mu$Jy beam$^{-1}$ for a 6 arcsecond beam. From top to bottom, we plot our low-powered, slow cluster simulation (Q36-v01-a25-C180), large, low-powered group simulation (Q36-v01-a25-G180), the high-powered, fast cluster simulation at 180 kpc (Q38-v98-a25-C180) and then at 20 kpc (Q38-v98-a25-C20). In the last row, we show the high-powered, fast group simulation (Q38-v98-a25-G180). We consider the last data output before the jet switches off (blue shaded region), 10 Myr into the remnant phase (red) and 50 Myr into the remnant phase (orange).}
    \label{fig:SB_dist_individual}
\end{figure*}

In Fig. \ref{fig:SB_dist_individual}, we plot the probability density function \footnote{The probability density function (PDF) of the surface brightness distribution is estimated using a kernel density estimator (KDE) applied to the sampled values of 
$I_{\nu}$ (in mJy arcsec$^{-2}$), providing a smooth representation of the underlying distribution.} of surface brightness (at 150 MHz) for five sources at three time steps $t_{\rm{on}}$ (in blue), $t_{\rm{on}} + 10$ Myr (orange), and  $t_{\rm{on}} + 50$ Myr (red). The left and right columns show the distributions for the same simulation at a redshift of 0.05 and 1, respectively. Old remnant radio sources ($t \sim t_{\rm{on}} + 50$ Myr) are generally characterised by narrower surface brightness distributions (above a realistic detectable limit) than active sources for $z=0.05$ and $1$. At low-redshift (left-hand column of Fig. \ref{fig:SB_dist_individual}), the largest spread in surface brightness values during the remnant phase is seen in low-powered cluster sources (the first row of Fig. \ref{fig:SB_dist_individual}) reflecting the slower fading rates of these sources. In most cases, more than 50\% of the still-emitting remnant lobe would be detectable above a noise level similar to that achievable by LOFAR (around $23 \mu$ Jy beam$^{-1}$ for an integration time of approximately 100 hours, \citealp{Tasse_2021}) across both redshifts. For a low-powered remnant in a group environment (second row of Fig. \ref{fig:SB_dist_individual}), 20-30\% of the source is undetectable after 50 Myrs at a low redshift, while at high redshift, around 70\% of the source would be undetectable.\\

For remnants created by high-powered progenitors, the upper limit of the surface brightness falls by as much as 70-85\% within the first 10 Myr. As shown in the bottom three rows of Fig. \ref{fig:gridactives}, the regions of highest surface brightness are concentrated at the jet channel and at the head of the lobes. A rapid narrowing in surface brightness distribution is consistent with a short ($\sim$ 10 Myr) fading time of emission from the jet and lobe head. \\

At higher redshift (right-hand column of Fig. \ref{fig:SB_dist_individual}), the spread in surface brightness values in low-powered sources is significantly narrower due to the prevalence of inverse-Compton scattering (see Fig. \ref{fig:loss_grid}). For the large, low-powered remnant in the group environment (right-column, second row down), the surface brightness distribution at $t_{\rm{on}} + 50$ Myr is similar to that at $t_{\rm{on}}$, suggesting that it takes $> 50$ Myr for the emission at 150 MHz to fade. Conversely, fading occurs rapidly for high-powered sources, and the surface distributions at $t_{\rm{on}} + 50$ Myr show the narrowest range of values. \\

\begin{figure*}
    \centering
    \includegraphics[width=0.90\linewidth]{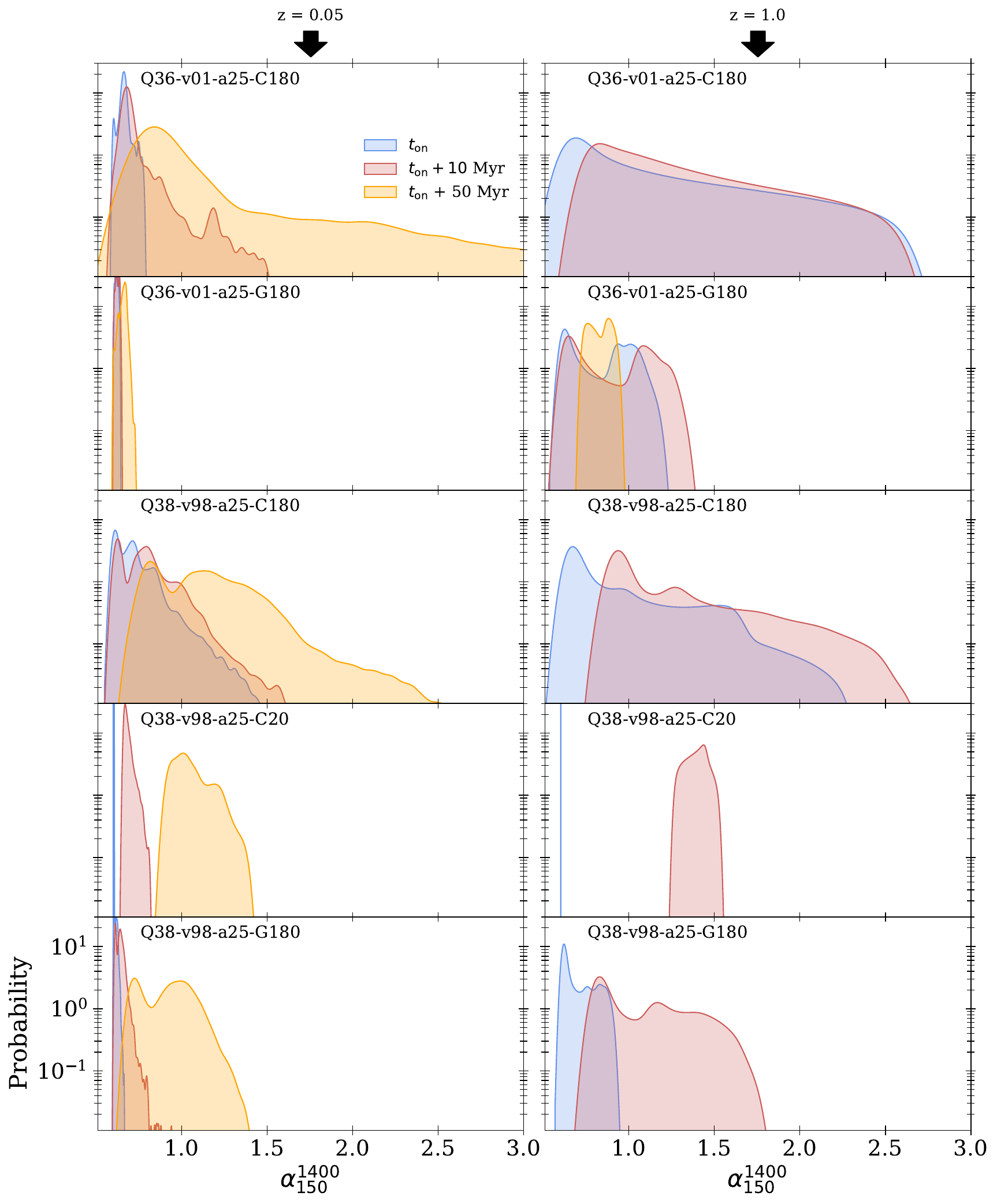}
    \caption{Analogous to Fig. \ref{fig:SB_dist_individual} but showing the probability density of spectral index between 150 MHz and 1400 MHz. For the distributions at $z = 1.0$, there is no detectable emission at 1400 MHz at $t_{\rm{on}} + 50$ Myr.}
    \label{fig:spec-index_dist_individual}
\end{figure*}

In Fig. \ref{fig:spec-index_dist_individual}, we consider the spatial distribution of $\alpha_{150}^{1400}$ for the same sources, and at the same time steps. Shorter active times generally lead to narrower ranges of spectral index when considering equivalent jet and environment initial conditions. The largest range in spectral index values at $t_{\rm{on}}$ is associated with extended sources in the cluster environment (first and third row). These are also the sources which show the largest spread of spectral indices during the remnant phase at both low and high redshift. The most uniform spectral indices are associated with compact sources and remnants from low-powered group sources at low redshift. Despite having the shortest active time, compact sources from high-powered progenitors show a larger spectral index range than their low-powered counterparts at late times. At $z=1$ (right-hand column of Fig. \ref{fig:spec-index_dist_individual}) rapid fading at 1400 MHz means no spectral index data is available past $10-15$ Myr, with remnants of high-powered progenitors fading faster than those of low-power.

\subsection{Core Prominence}
\label{sec:core_prom}



The core prominence (CP) is a selection criterion used in observational studies of radio remnants \citep[e.g.][]{Brienza_2017, Jurlin_2021, Quici_2021}. It describes the relative brightness of a radio core to its extended emission. As the nuclear emission greatly reduces at the start of the remnant phase, the CP values of remnant sources are expected to fall to zero. In the literature, authors calculate the CP of remnant candidates as the ratio of core flux density to flux density of extended emission \citep[e.g.][]{Brienza_2017, Jurlin_2021}, $CP = S_{\rm{core}} / S_{\rm{extended}}$.\\

A few works have reported CP values for samples of candidate remnants. \cite{Brienza_2017} find an average CP value of $3.3\times10^{-4}$, and in \cite{Jurlin_2021}, the authors find a higher range of CP values within $0.0019 - 0.15$ for 17 of their candidate remnants. 
\\

In this work, we take the core flux densities at 6000 MHz, while the flux densities of the extended emission are measured at 1400 MHz. This is similar to the process of \cite{Jurlin_2021} where the Very Large Array (VLA) is used to image the core at 6000 MHz and the extended 1400 MHz emission is taken from NRAO VLA Sky Survey (NVSS) images. To determine the flux density in the core region, we consider the brightest pixel within a region of radius $2$ kpc (corresponding to $2$ arcseconds at $z=0.05$ and $0.2$ arcseconds at $z=1$) around the origin of our grid. The extended flux density is calculated as the total flux density of all emitting material with the flux density from the core removed, $S_{\rm{extended, \ 1400 MHz}} = S_{\rm{total, \ 1400 MHz}} - S_{\rm{core, \ 1400 MHz}}$.  Since we inject our jet fluid through an injection region placed on the computational grid and hence do not simulate the active nucleus and base of the jet below the injection radius of 0.75 kpc, we choose to only consider the remnant phase when discussing CP evolution. Further, the CP values calculated here are lower limits since we do not include the VLBI core.\\ 

\begin{figure*}
    \centering
    \includegraphics[width=0.84\linewidth]{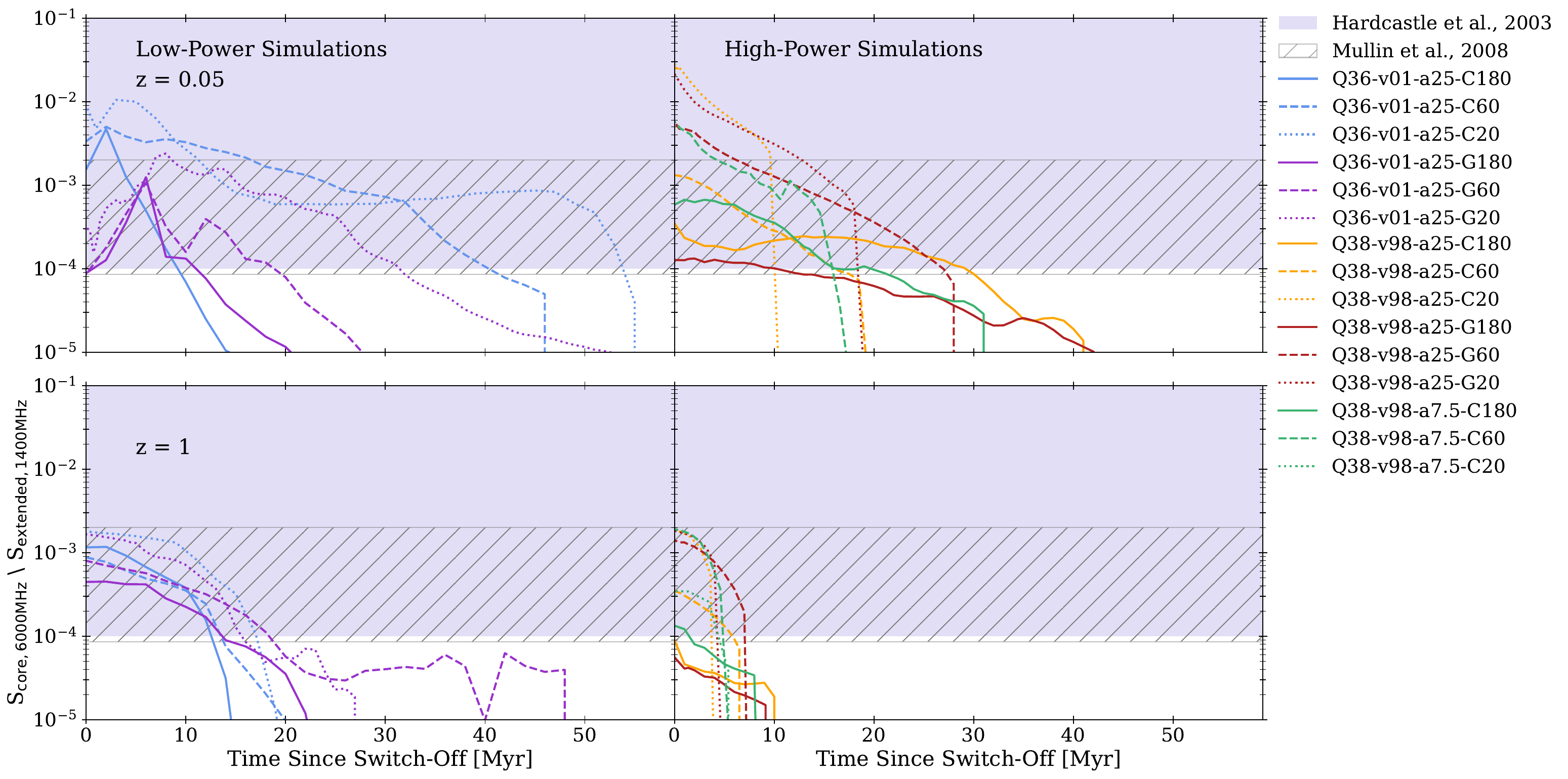}
    \caption{The evolution of core prominence (CP) values following the time of switch-off for all sources at a redshift of 0.05 (top row) and 1 (bottom row). We show the ranges of CP values obtained by \cite{Hardcastle_2003} using the B2 bright sample (shaded purple region), and by \cite{Mullin_2008} using a subsample of the 3CR survey (hashed grey region).
    }
    \label{fig:multi_redshift_core_prom}
\end{figure*}

We plot the temporal evolution of core prominence in the remnant phase for all sources in Fig. \ref{fig:multi_redshift_core_prom}. We compare the tracks of our simulations with the range of CP values obtained for active sources from the B2 bright sample (\citealp{Hardcastle_2003}, shaded blue region) and sources from the 3CR survey (\citealp{Mullin_2008}, hashed region). During the remnant stage, the CP values decrease (more rapidly at a higher redshift) as the 6000 MHz emission fades faster than the 1400 MHz emission. At higher redshifts, CP values are typically lower by a few dex as the core emission at 6000 MHz fades more rapidly than the extended emission at 1400 MHz. More rapid fading also generally yields a shorter time frame over which CP values are measurable at higher redshifts. We find that most remnant objects reach CP values below $10^{-4}$ at some point after switch-off; however, this does not occur immediately in the majority of cases. For larger sources, we generally find lower CP values at the start of the remnant phase compared with more compact sources, where the residual core-region emission is more significant. At all redshifts, residual lobe emission persists within the core region, preventing the CP value from dropping below $10^{-4}$ immediately after switch-off for the considered core size (radius of 2 arcseconds at $z=0.05$ and 0.2 arcseconds at $z=1$). This may have implications for remnant classification if the time it takes plasma to exit the core region delays the drop of CP to zero after jet switch-off. In some cases (namely our simulations in the cluster where plasma is more confined by a higher central density and pressure), we find residual plasma keeps the CP value of our simulated remnants above $10^{-4}$ for 30-40 Myr after switch-off. \\

\section{Remnant Selection Timescales}
\label{sec:remnant selection timescales}
\begin{figure*}
    \centering
    \includegraphics[width=0.95\linewidth]{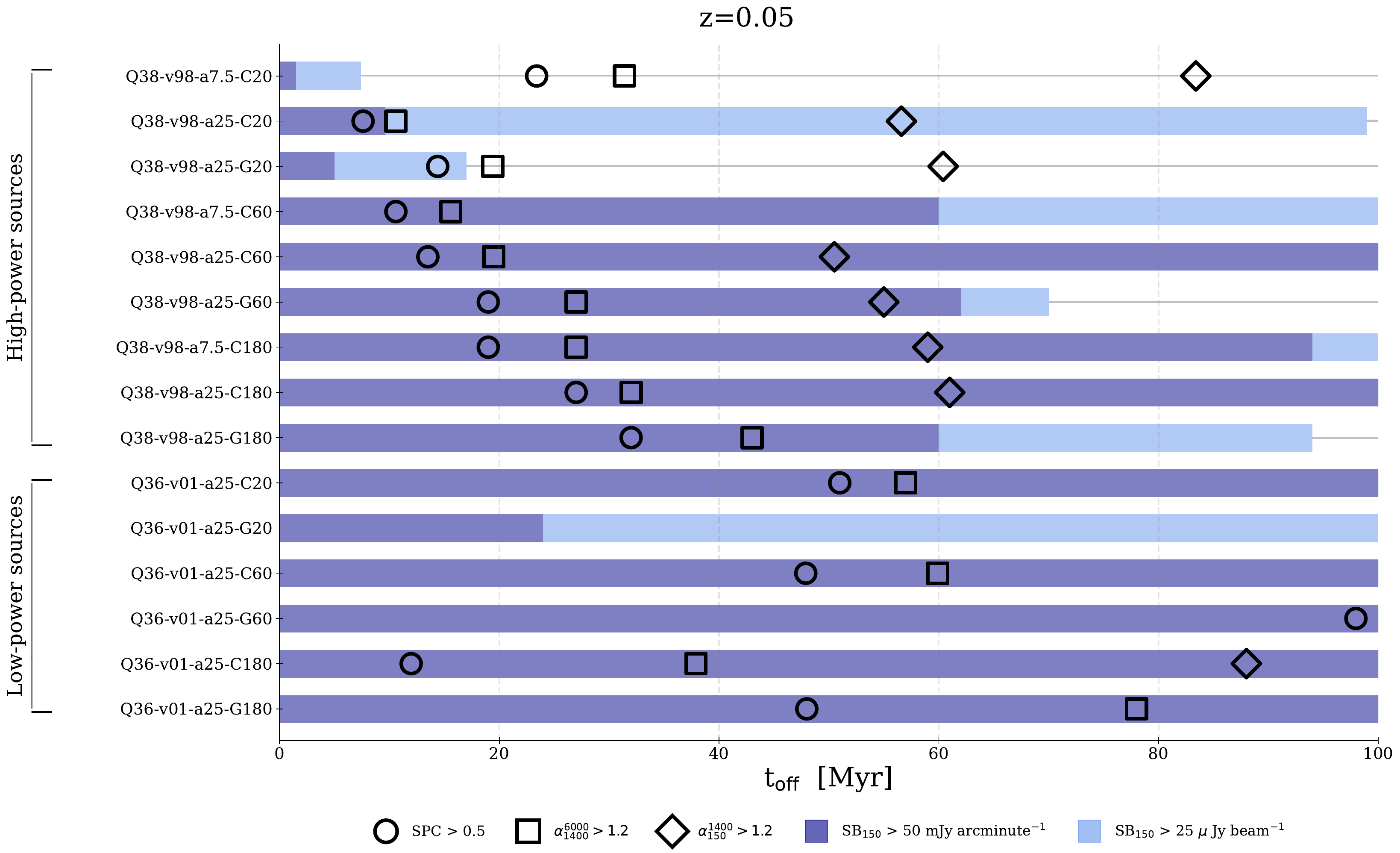}
    \caption{A timeline showing when different spectral thresholds are crossed at $z=0.05$ as a function of time since switch-off. Markers indicate when the integrated spectral curvature exceeds 0.5 (circles), when the spectrum is ultra-steep between 6000 and 1400 MHz (squares), and when the spectral index becomes ultra-steep between 150 and 1400 MHz (diamonds). The purple shaded regions denote where the median surface brightness of the source is above 50 mJy arcminute$^{-2}$ while blue regions indicate brightness above 25 $\mu$Jy beam$^{-1}$, comparable to the LOFAR rms noise value given in \cite{Jurlin_2021}.}
    \label{fig:timeline_plot_z=0.05}
\end{figure*}

\begin{figure*}
    \centering
    \includegraphics[width=0.95\linewidth]{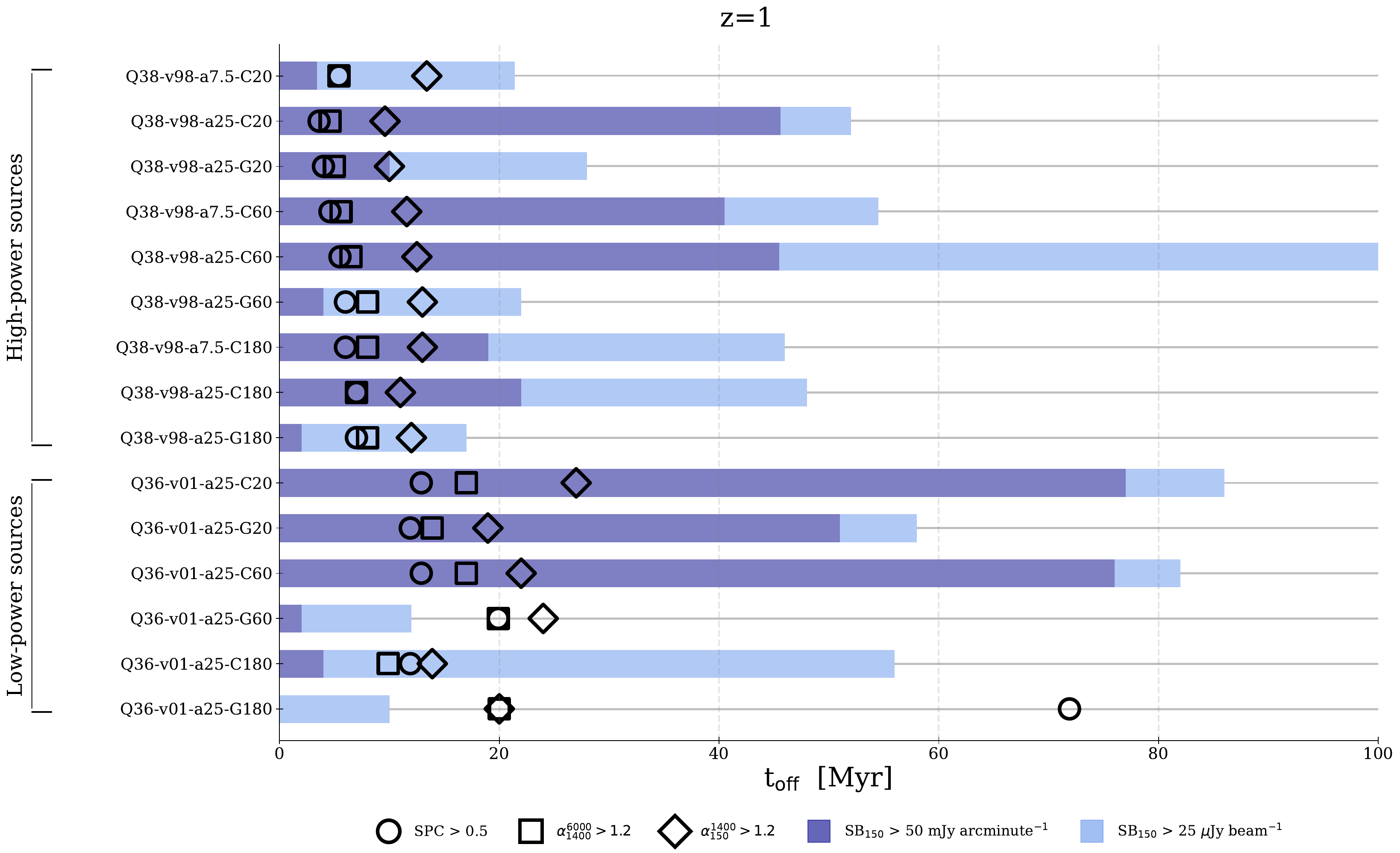}
    \caption{Analogous to Fig.\ref{fig:timeline_plot_z=0.05} but with the spectral properties processed at a higher redshift of 1.0. }
    \label{fig:timeline_plot_2_z=1}
\end{figure*}

In this section, we summarise the timescales over which our simulated remnants develop the key spectral signatures often used in observational remnant identification studies. In Figs.\ref{fig:timeline_plot_z=0.05} and \ref{fig:timeline_plot_2_z=1} we show two timeline plots at redshifts of 0.05 and 1, respectively. On each plot, we indicate when our sources develop an ultra-steep spectral index between 150 and 1400 MHz ($\alpha_{150}^{1400} > 1.2$) as well as between 1400 and 6000 MHz ($\alpha_{1400}^{6000} > 1.2$), strong spectral curvature ($\alpha_{1400}^{6000} - \alpha_{150}^{1400} > 0.5$), and a low median surface brightness value (SB$_{150 \rm{ MHz}} > 50$ mJy  arcminute$^{-2}$). We also estimate for how long each remnant would be visible to modern, low frequency instruments such as LOFAR. For this, we consider the duration for which the median surface brightness value is above a threshold of 25$\mu$Jy beam$^{-1}$ for a 5 arcsecond beam, similar to the reported rms value in \cite{Jurlin_2021}.\\

At low redshift (Fig. \ref{fig:timeline_plot_z=0.05}), we see a consistent sequence in the spectral evolution of our simulated remnants; the development of strong spectral curvature followed by spectral steepening between 6000 and 1400 MHz, and finally spectral steepening at the low-frequency end between 150 and 1400 MHz. This result is consistent with preferential fading of higher frequency emission. Typically, for remnants with high-power progenitors, spectral steepening occurs most rapidly for compact objects. Conversely, larger remnants with low-power progenitors will develop steeper and more curved spectra at a faster rate than compact ones. As reported in Section \ref{sec:lowzremnants} and shown in Fig. \ref{fig:timeline_plot_z=0.05}, one high-power, large remnant (Q38-v98-a25-G180) does not develop an ultra steep, low frequency spectral index within the simulated time. Furthermore, remnants run in the group environment will develop spectral features slower than the equivalent cluster simulation. Except for compact sources, all low-redshift remnants will display a curved spectrum and ultra-steep spectral index between 1400 and 6000 MHz before the median surface brightness falls below 50 mJy arcminute$^{-2}$ at 150 MHz and long before the source falls below the 25$\mu$Jy beam$^{-1}$ threshold at 150 MHz. \\

As redshift increases (Fig. \ref{fig:timeline_plot_2_z=1}), the timescales for spectral features to develop shorten. The trend for compact remnants with high-power progenitors to develop steep spectral features before larger ones is largely preserved at $z=1$ as is the spectral sequence of strong spectral curvature, followed by an ultra steep high-frequency spectral index and then an ultra steep low-frequency spectral index. Remnants in the group environment (which are often close to, or already below, the 50 mJy arcminute$^{-2}$ low surface brightness threshold at the end of the active phase) remain visible for less than half of the duration of the equivalent cluster simulation. For low-powered, group simulations, this can mean that strong spectral curvature and steep spectral indices are seen after the source has fallen below the rms noise threshold of $25 \mu$Jy beam$^{-1}$.\\

These results suggest that a candidate remnant radio galaxy observed with a spectral curvature larger than 0.5 and without a low-frequency spectral index less than 1.2 is likely to be younger than one with $\alpha_{150}^{1400} > 1.2$. Conversely, we consistently find that remnants with low-frequency spectral indices steeper than 1.2 indicate significantly aged plasma, particularly at low-redshift. The spectral curvature histograms presented in Fig. \ref{fig:speccurve_grid_all_sims} show lower counts for spectral curvature values above 0.5 than for values below 0.5. This suggests that each simulation spends a short amount of time with a spectral curvature above 0.5 before the source has faded at higher frequencies. Hence, the oldest remnant plasma seen in our simulations will have an ultra steep spectral index without significant spectral curvature. \\

\section{Discussion}
\label{sec:Discussion}

\subsection{Remnants and Their Environments}
\label{subsec:Remnants and their environments}

Some authors have argued that remnants would be observed preferentially in cluster environments, where the denser ambient medium would confine the ageing plasma for longer, reducing adiabatic losses \citep[e.g.][]{Murgia_2011} and increasing the visibility time of the remnant. In recent work, however, there is little supporting evidence (other than \citealp{White_inprep}) that the observed candidate remnant radio galaxy population are preferentially hosted by cluster environments. For example, \cite{Jurlin_2021} find only three out of their sample of 13 candidate remnants to reside in cluster environments based on optical host information, while \cite{Singh_2021} similarly find two of their sample of 15 candidates. A possible explanation for this is given in \cite{Turner_2026}, where the dense cluster environment may cause some remnant lobes to implode. From our simulations, we can also comment on possible reasons why this is the case. \\

Firstly, our results suggest that selection of remnants using a low surface brightness criterion would select remnants in the case where the remnant is in a group-like environment and/or where the progenitor has a higher jet power, which fade faster than low-powered sources \citep{Turner_2018}. We have shown that sources run in the group environment would satisfy the low surface brightness criterion before an equivalent simulation in the cluster environment (see Figs. \ref{fig:timeline_plot_z=0.05} and \ref{fig:timeline_plot_2_z=1}). This may bias remnant selection toward objects in group environments, especially if nuclear activity restarts after some time. At low redshift, remnants in groups are more likely (compared to those in clusters) to fade below a low surface brightness cutoff before jet production is reignited in the active nucleus. In some cases (such as remnants of low-power progenitors at higher redshifts), this low-surface brightness emission can be present without strong spectral curvature or an ultra-steep, low-frequency spectral index. Furthermore, selection based on low-surface brightness at higher redshifts may misidentify active sources as remnants. Comparing the timelines shown in Fig. \ref{fig:timeline_plot_z=0.05} and Fig. \ref{fig:timeline_plot_2_z=1} show one instance where the $\rm{SB}_{150} < 50 \ mJy \ arcminute^{-1}$ is reached before the source is switched off (Q36-v01-a25-G180). Several other sources would meet the low surface brightness within a few Myr after switch-off (e.g. Q38-v98-a25-G180 and Q36-v01-a25-G60).\\

Secondly, we suggest that remnant selection using the amorphous morphology criteria is biased towards remnants of high-power progenitors and/or those in group environments. The amorphous appearance of radio plasma is taken to be an indicator of remnant nature in several works \citep{Brienza_2016, Brienza_2017, Mahatma_2018}. The development of amorphous structures is difficult to quantify; however, remnants in the group environment show a greater departure from their active morphology 50 Myr after the jets have turned off. In \PaperII ~, we established that simulations in this environment can remain overpressured in the remnant phase for significantly longer than the equivalent cluster model. This overpressure appears to drive the transition to an amorphous morphology more so than the onset of buoyancy. The most rapid change in morphology is seen in FR-II sources, where the overpressure is larger, and the bulk fluid motions result in greater mixing with the environment.

\section{Conclusions}
\label{sec:conclusions}

In this paper, we have used three-dimensional, numerical, hydrodynamical simulations to study the morphological and spectral evolution of active and remnant RLAGN. We have sampled a parameter space that covers low ($10^{36}$ W), and high ($10^{38}$ W) jet powers, strongly relativistic (0.98c) and subrelativistic (0.01c) injection velocities, and a wide range of active lifetimes ($0.5 - 100$ Myr) to generate a representative sample of simulated radio sources. These sources propagate through group and cluster-like environments derived from cosmological simulations. In post-processing, we have explored the radio emission at a redshift range of $0.05 \leq z \leq 1.0$ and within a frequency range of $150 \rm{\ MHz} \leq$ $\nu$ $\leq 6000 \rm{\ MHz}$. We find the following:\\

(i) Simulations in the group environment are consistently dimmer than their cluster counterparts and fall below the low-surface brightness threshold (SB$_{150}$ < 50 mJy arcminute$^{-2}$) earlier in their evolution. Remnant selection methods that take low-surface brightness into account may, therefore, be biased towards remnants not hosted by cluster environments. Further, we confirm that the surface brightness is redshift-dependent and decreases systematically with increasing redshift. Redshift effects should be considered when interpreting the surface brightness of remnant samples.\\

(ii) Amorphous morphologies develop at a faster rate for remnants of  high-powered progenitors than for those of low-power progenitors. Selection methods that take amorphous morphologies into account may be biased towards remnants from FR-II progenitors.\\

(iii) Remnant radio sources have consistently more uniform surface brightness distributions than active sources, which reflects a rapid loss of compact bright features, particularly in remnants of high-powered progenitors.\\

(iv) We find a significant overlap in spectral index values between our remnant and active sources. Although no active sources have an ultra-steep ($\alpha_{150}^{1400} > 1.2$) spectral index, each  remnant simulation spends a considerable amount of time (often longer than the preceding active phase) with low-frequency spectral indices that overlap with the active population. \\

(v) Once the jet switches off, spectral steepening occurs most rapidly in remnants of high-powered progenitors within cluster environments. This is in keeping with the behaviour in the active phase. Low-powered sources also steepen more slowly in group environments during both the active and remnant phases.\\

(vi) The majority of simulated remnants have core prominence values below $10^{-4}$ at some point after switch-off. However, we find that residual radiating material at the base of the lobe prevents the CP value from dropping below $10^{-4}$ immediately following switch-off. At low redshift, this is more prominent in cluster simulations where the higher density and pressure confines the plasma to the core region for longer, increasing the plasma crossing time out of the core region.\\
 
(vii) We confirm that the time for radiating particles to age and spectral features to develop shortens at higher redshift, driven by the increased significance of inverse-Compton scattering. For low-powered sources in the group environment, spectral steepening is driven by inverse-Compton scattering regardless of redshift. For high-powered sources in the cluster environment, synchrotron loss processes dominate spectral ageing and can still be significant at high redshift.\\

Overall, our results highlight the effects of varying jet and environment properties on the morphological and spectral evolution of remnant radio galaxies. We estimate that instruments such as LOFAR could currently detect approximately 70–80\% of the emission from our faintest, 50 Myr-old sources at low redshift in low-density, group-like environments. Our analysis of common remnant selection criteria aims to improve the accurate identification of large samples of radio remnants, which are essential for constraining jet populations and understanding feedback over the entire AGN lifecycle. This effort is particularly timely with the upcoming SKA observations, which are expected to significantly expand the observed populations of remnant radio sources. Developing a more comprehensive theoretical framework for remnant evolution will require simulations that self-consistently model magnetic fields and incorporate more dynamic environments. A detailed comparison of the spectral properties of magnetohydrodynamical simulations of AGN remnants will be carried out in future work.

\section{Acknowledgements}

We thank the anonymous referee whose comments have added value to the manuscript. GSCS thanks the Australian Government for an Australian Government Research Training Program RTP scholarship and the CSIRO for an ATNF studentship. SS acknowledges the Australian Research Council grant DP240102970. This research was carried out using high-performance computing infrastructure provided by the Tasmanian Partnership for Advanced Computing (TPAC). We acknowledge the support of the TPAC team, in particular Geli Kourakis and John Miezitis, for providing timely and useful troubleshooting support. 


\bibliography{references}


\end{document}